\documentclass[twocolumn,showpacs,showkeys,preprintnumbers]{revtex4}

\usepackage{graphicx}

\def\spose#1{\hbox to 0pt{#1\hss}}
\def\simlt{\mathrel{\spose{\lower 3pt\hbox{$\mathchar"218$}}
     \raise 2.0pt\hbox{$\mathchar"13C$}}}
\def\simgt{\mathrel{\spose{\lower 3pt\hbox{$\mathchar"218$}}
     \raise 2.0pt\hbox{$\mathchar"13E$}}}
\def\simpropto{\mathrel{\spose{\lower 3pt\hbox{$\mathchar"218$}}
     \raise 2.0pt\hbox{$\propto$}}}
\newcommand\lsim{\mathrel{\rlap{\lower4pt\hbox{\hskip1pt$\sim$}}
        \raise1pt\hbox{$<$}}}
\newcommand\gsim{\mathrel{\rlap{\lower4pt\hbox{\hskip1pt$\sim$}}
        \raise1pt\hbox{$>$}}}

\def\ie{{\it i.e. }}
\def\eg{{\it e.g. }}

\def\rmb{{\rm b}}

\def\rme{{\rm e}}

\def\rmm{{\rm m}}
\def\rmp{{\rm p}}
\def\rmr{{\rm r}}
\def\rmv{{\rm v}}
\def\bfa{{\bf a}}
\def\bfb{{\bf b}}
\def\bfc{{\bf c}}

\def\bfg{{\bf g}}

\def\bfr{{\bf r}}
\def\bfv{{\bf v}}
\def\bfx{{\bf x}}
\def\bfA{{\bf A}}
\def\bfB{{\bf B}}

\def\bfD{{\bf D}}

\def\bfH{{\bf H}}
\def\bfI{{\bf I}}

\def\bfN{{\bf N}}
\def\bfQ{{\bf Q}}
\def\bfP{{\bf P}}

\def\bfT{{\bf T}}
\def\bfU{{\bf U}}

\def\calO{{\cal O}}
\def\calD{{\cal D}}

\def\bfvb{{\bf v}_{\rm b}}

\def\hatbfn{{\bf \hat{n}}}
\def\hatbfc{{\bf \hat{c}}}
\def\hatbfp{{\bf \hat{p}}}

\def\ln{{\rm ln}}
\def\exp{{\rm exp}}
\def\diag{{\rm diag}}
\def\lim{{\rm lim}}
 
\def\grad{{\bf\nabla}}
\def\Tr{{\rm Tr}}
\def\kB{{k_{\rm B}}}
\def\Tb{{T_{\rm b}}}
\def\Tc{{T_{\rm c}}}
\def\TRJ{{T_{\rm RJ}}}
\def\nBB{{n^{\rm BB}}}
\def\dnSZ{{\Delta n^{\rm SZ}}}
\def\dn#1{{\Delta n^{(#1)}}}

\def\Te{{T_{\rm e}}}
\def\me{{m_{\rm e}}}
\def\mH{{m_{\rm H}}}
\def\ve{{v_{\rm e}}}
\def\nel{{n_{\rm el}}}
\def\sigmaT{{\sigma_{\rm T}}}
\def\Thetae{{\Theta_{\rm e}}}
\def\zc{{z_{\rm c}}}
\def\overleftrightarrow#1{\vbox{\ialign{$##$\cr
 {\leftarrow}\mkern-6mu\cleaders\hbox{$#1$}\hfill
 \mkern-6mu{\to}\cr\noalign{\kern-.2326ex\nointerlineskip}\hfil#1\hfil\cr}}}

\def\Dt{\overleftrightarrow{\bfD}}
\def\Ht#1{\overleftrightarrow{\bfH}_{(#1)}}
\def\It{\overleftrightarrow{\bfI}}
\def\Nt{\overleftrightarrow{\bfN}}
\def\Pt{\overleftrightarrow{\bfP}}
\def\Qt{\overleftrightarrow{\bfQ}}
\def\Tt{\overleftrightarrow{\bfT}}
\def\Ut#1{\overleftrightarrow{\bfU}_{(#1)}}
\def\pvector#1{{\hatbfp^{(#1)}}}
\def\SigmaOp{{\hat{\Sigma}\star}}
\def\AngleAverage#1{{\left\langle#1\right\rangle_{\hatbfc'}}}
\def\uv{{\overline{u_\rmv}}}
\def\udT{{\overline{u_{\rm\Delta T}}}}
\def\ySZ{{\overline{y_{\rm SZ}}}}
\def\zrei{{z_{\rm rei}}}
\def\tauobs{{\tau_{\rm obs}}}
                                    
\begin{document}

\title{CMB Spectral Distortions from the Scattering of Temperature
Anisotropies}
\author{Albert Stebbins$^{1}$}
\affiliation{$^1$ Center for Particle Astrophysics,
Fermi National Accelerator Laboratory, Batavia, IL 60510}
\email{stebbins@fnal.gov}

\date{\today}

\begin{abstract}
Thomson scattering of CMBR temperature anisotropies will cause the spectrum of
the CMBR to differ from blackbody even when one resolves all anisotropies.  A
formalism for computing the anisotropic and inhomogeneous spectral distortions
of intensity and polarization is derived in terms of Lorentz invariant central
moments of the temperature distribution.  The formalism is non-perturbative,
requiring neither small anisotropies nor small metric or matter
inhomogeneities; but it does assume cold electrons.  The low order moments are
not coupled to the higher order moments allowing one to truncate the equations
without any loss of accuracy.  This formalism is applied to a standard
$\Lambda$-CDM cosmology after reionization where the temperature anisotropies
are dominated by the Doppler effect for the bulk motion of the gas with respect
to the CMBR frame.  The resultant spectral distortion is parameterized by
$u\approx3\times10^{-8}$, where in this case $u$ is observationally degenerate
with the Sunyaev-Zel'dovich (SZ) $y$ parameter. In comparison the expected
thermal SZ $y$-distortion from the hot IGM is expected to be $\simgt30$ times
larger. However at $z\simgt5$ the effect described here would have been the
dominant source of spectral distortions.  The effect could be much larger in
non-standard cosmologies.
\end{abstract}


\preprint{FERMILAB-PUB-07-065-A}
\maketitle

\section{Introduction}
\label{intro}

The cosmic microwave background radiation (CMBR) is observed to have a spectrum
extremely close to a blackbody (a.k.a. thermal or Planckian) spectrum with
temperature $T=2.725\,$K \cite{firas94a,firas94b,firas96,firas02}. In addition
to contamination from foreground radio and far-IR sources, deviations from a
thermal spectrum is observed in the direction of concentrations of hot gas
(galaxy clusters) due to the thermal Sunyaev-Zel'dovich effect (tSZ)
\cite{SZreview}.  If one had better sensitivity one should see some amount of
tSZ spectral distortions everywhere since one expect there is ionized gas with
$T_{\rm g}\simgt10^6$\,K along every line-of-sight.

In the Earth frame the brightness temperature of the CMBR is observed to vary
by several mK in a dipole pattern; however this would disappear if the observer
were boosted into the "CMBR frame".  After accounting for this velocity dipole
there is also observed residual primary anisotropies at the 10s of $\mu$K level
concentrated on the sub-degree scales.  These anisotropies are expected and
observed to be very close to blackbody \ie the temperature varies with
direction on the sky but the spectrum in each direction on the sky is close to
a blackbody.  If one observes the CMBR with a finite beam instrument, because
of the anisotropies, and because the average of a blackbody spectra with
different temperature is not a blackbody, the observed spectrum will exhibit a
spectral distortion from blackbody \cite{cs04}.  This "beam mixing distortion"
is very similar to, and can be confused with, the tSZ distortions mentioned
above.  However this is {\it not} the topic of this paper.
 
Here the mixing of anisotropic blackbody spectra due to scattering of the
radiation is examined.  These distortions will occur even with arbitrarily fine
angular resolution; and thus are called "resolved spectral distortions" in
contrast to the "unresolved spectral distortions" from beam mixing.  As we
shall see, apart from a small and isotropic primordial tSZ effect, the
anisotropic blackbody approximation is correct to 1st order in the amplitude of
inhomogeneities in the universe, and the resolved distortions caused by
coupling of scattering and anisotropies arise in 2nd order.  Just from this
consideration one can expect the spectral distortion to be very small.

\subsection{Roadmap}
\label{roadmap}

Here is a brief description of what is included in this paper
\begin{description}
\item{\bf\S\ref{sec:Spectrum:transform}:} Temperature transform representation.
\item{\bf\S\ref{sec:Spectrum:moments}:} Decompose transform into moments.
\item{\bf\S\ref{app:MomentInverse}:} Practical method to compute moments.
\item{\bf\S\ref{sec:Spectrum:Lorentz}:} Central moments shown to be Lorentz
invariant.
\item{\bf\S\ref{sec:Spectrum:FokkerPlanck}:} Moments are coefficients in
Fokker-Planck expansion.
\item{\bf\S\ref{sec:Spectrum:Asymptotes}:} Some properties of expansion.
\item{\bf\S\ref{sec:Spectrum:PowerMoments}:} Relation to commonly used
power law moments.
\item{\bf\S\ref{sec:Spectrum:SZ}:} tSZ $y$-distortion in terms of moments.
\item{\bf\S\ref{app:Frames}:} Global frames: 3+1 description of space-times.
\item{\bf\S\ref{sec:Boltzmann:redshift}:} Redshifting of photons in terms of
global frame velocity gradients.
\item{\bf\S\ref{sec:Boltzmann:Thomson}:} Thomson cross-section as a
convolution.
\item{\bf\S\ref{sec:Boltzmann:Thomson}:} Apply convolution to the temperature
transform.
\item{\bf\S\ref{sec:Boltzmann:Temperature}:} Apply convolution to moments.
\item{\bf\S\ref{sec:SpectralDynamics:Anisotropies}:} Spectral distortions
are driven by temperature anisotropies.
\item{\bf\S\ref{sec:SpectralDynamics:Origin}:} Cosmological initial conditions
and origin of spectral distortions. 
\item{\bf\S\ref{sec:SpectralDynamics:Perturb}:} Order in cosmological
perturbation theory where different terms become non-zero.
\item{\bf\S\ref{app:Polarization}:} Generalizes all of the above to polarized
light.
\item{\bf\S\ref{sec:SpectralDynamics:DeltaToverT}:} Relates $\bar{T}$ to usual
$\Delta T/T$ language.
\item{\bf\S\ref{sec:SpectralDynamics:Lowest}:} Lowest order equations for small
perturbations.
\item{\bf\S\ref{sec:SpectralDynamics:Single}:} All moments for small optical
depth.
\item{\bf\S\ref{sec:SpectralDynamics:tSZ}:} Adds tSZ $y$-distortions.
\item{\bf\S\ref{sec:uAverage}:} Application to expected mean spectral
distortion.
\item{\bf\S\ref{app:ComputingVelocities}:} Details for computing mean
distortion.
\item{\bf\S\ref{sec:Discussion}:} Discussion of results.
\end{description}
Many of these sections may be skipped as technical details, depending on your
interests.

\section{Representation of the Spectrum}
\label{sec:Spectrum}

Usually one quantifies the flux of photons by a brightness\footnote{N.B.
$[\cdots]$, not $(\cdots)$, are used to indicate functional dependencies.  Also
$\cdots$ are used as a shorthand to replace arguments which, in the context of
a particular equation, are unimportant.}  $I_\nu[\hatbfc,\bfx,t]$ where $\nu$,
$\bfx$, $t$ and $\hatbfc$ are respectively the frequency, spatial position,
time, and direction in which the photons are traveling (N.B. usually one uses
$\hatbfn=-\hatbfc$).  The quantities we will end up dealing with are a highly
transformed form of $I_\nu$.  The first step is to transform to the
dimensionless quantum mechanical occupation number $n[\nu,\cdots]\equiv c^2
I_\nu[\cdots]/(4\pi\hbar\nu^3)$ where $h$ is the Planck constant.  Implicit in
these definitions is a choice of rest-frame.  Of course $\nu$ will be Doppler
shifted from one rest-frame to another, but the quantity $n$ is rest-frame
independent (\ie a Lorentz invariant) once one takes into account the Doppler
shifting of the argument $\nu$.

Other common representations of the spectra is the {\it brightness
temperature}, $\Tb$, defined by $\Tb\equiv h\nu/(\kB\,\ln[1+1/n])$, which gives
the temperature of a blackbody which would produce the equivalent brightness at
that frequency. Here $\kB$ is the Boltzmann constant.  If $n\gg1$ this reduce
to the {\it Rayleigh-Jeans temperature}, $\TRJ=h\nu\,n/\kB$.  Both $\Tb$ and
$\TRJ$ are frame-dependent.

\subsection{The Temperature Transform}
\label{sec:Spectrum:transform}

A blackbody spectrum has frequency dependence
\begin{equation}
n[\nu,\cdots]=\nBB\left[{h\nu\over\kB\,T[\cdots]}\right] \qquad
\nBB[x]={1\over e^x-1}\ .
\label{BBapprox}
\end{equation}
where $T$ is the temperature.  If for a given $\bfx$, $t$, $\hatbfc$, the
frequency dependence differs from this functional form we say there is a {\it
spectral distortion}.

Next define the {\it temperature transform} \cite{sz72,cj75,salas92,cs04},
$q[\ln[T]\,\cdots]$, of $n[\nu,\cdots]$, defined implicitly by
\begin{equation}
n[\nu,\cdots]=\int_{-\infty}^\infty d\ln[T]\,q[\ln[T],\cdots]
                                      \,\nBB\left[{h\nu\over\kB\,T}\right]\ .
\label{temperaturetransform}
\end{equation}
In \S2 it is shown that $q[\ln[T],\cdots]$ is sufficient to represent the
spectral distortions that we are interested in here.

\subsection{Logarithmic and Central Moments}
\label{sec:Spectrum:moments}

It is useful to characterize $q[\ln[T],\cdots]$ by it's {\it logarithmic
moments}
\begin{equation}
\eta_{(n)}[\cdots]\equiv
\int_{-\infty}^\infty d\ln[T]\,\ln[T]^n\,q[\ln[T],\cdots]
\label{logmoments}
\end{equation}
It is shown how to compute $\eta_{(n)}$ from $n[\nu,\cdots]$ in
appendix~\ref{app:MomentInverse}. From these moment we will define other
parameters beginning with the {\it grayness parameter}
\begin{equation}
\label{grayness}
g\equiv 1-\eta_{(0)}\ .
\end{equation}
If $\eta_{(0)}\ne0$ then $q[\ln[T],\cdots]/\eta_{(0)}$ integrates to unity and
can be thought of as a probability distribution function (pdf) for
$\ln[T]$. Using this pdf one can define the average
\begin{equation}
\langle f[T,\cdots]\rangle\equiv
\int_{-\infty}^{\infty} d\ln[T]\,{q[\ln[T],\cdots]\over \eta_{(0)}}\,f[T,\cdots]
\end{equation}
so that $\eta_{(n)}=\eta_{(0)}\,\langle\ln[T]^n\rangle$.  Define the {\it mean
logarithmic temperature} as
\begin{equation}
\label{meantemp}
\bar{T}\equiv e^{\langle \ln[T]\rangle}=e^{\eta_{(1)}/\eta_{(0)}}
,\end{equation}
and the central moments
\begin{equation}
u_{(n)}\equiv\left\langle\ln\left[{T\over\bar{T}}\right]^n\right\rangle
=\sum_{k=0}^n {(-1)^kn!\over k!\,(n-k)!}\,
{\eta_{(k)}\over \eta_{(0)}}\,\left({\eta_{(1)}\over \eta_{(0)}}\right)^{n-k}
\label{centralmoments}
.\end{equation}
Note that by definition $u_{(0)}=1$ and $u_{(1)}=0$.  The most important
central moment is the 1st non-trivial one, $u_{(2)}$, and for this 
reason we use the special notation
\begin{equation}
u\equiv u_{(2)}
\end{equation}
and give $u$ the name: {\it width of the temperature distribution}.  A graybody
spectrum has $u_{(n)}=0$ for $n>0$ and $g>0$ and this is why $g$ got the name
``grayness''.

\subsection{Lorentz Transformations}
\label{sec:Spectrum:Lorentz}

Under a Lorentz transformation, from frame 1 to 2, the various quantities are
Doppler shifted
\begin{eqnarray}
      \nu_2       &=&    {\nu_1\over1+z_{12}}         \\ \nonumber
  \bar{T}_2       &=&{\bar{T}_1\over1+z_{12}}         \\ \nonumber
  n_2[\nu,\cdots] &=&n_1[(1+z_{12})\,\nu,\cdots] \\ \nonumber
q_2[\ln[T],\cdots]&=&q_1[\ln[T]+\ln[1+z_{12}],\cdots]
\end{eqnarray}
The parameters $g$ and $u_{(n)}$ are Lorentz invariant, and this is the main
reason why it is convenient to parameterize the spectral distortion by $g$ and
the $u_{(n)}$.

\subsection{Fokker Planck Expansion}
\label{sec:Spectrum:FokkerPlanck}

Another reason why the central moments are useful is that the temperature
transform of the background radiation spectra will generically give a very
narrow distribution of temperature, corresponding to a small overall spectral
distortion.  That being the case, to the extent that the blackbody spectrum
$\nBB$ is well represented by the Taylor series expansion of it's argument,
$\nu/T$, then the spectral distortion is given by the moments of the
temperature distribution.  A mnemonic representation of this is to imagine 
expanding $q$ as a sum of derivatives of Dirac $\delta$-functions centered
on some fiducial temperature $T_0$, \ie
\begin{equation}
q[\ln[T],\cdots]=\sum_{m=0}^\infty{(-1)^m\over m!}\,d_{(m)}[\cdots]\,
{d^m\delta\left[\ln\left[{T\over T_0}\right]\right]\over d\ln[T]^m}\ .
\label{DeltaExpansion}
\end{equation}
Substituting this into the temperature transform one finds, by integrating by
parts and assuming $\lim_{\ln[T]\rightarrow\pm\infty}q=0$, that
\begin{equation}
n[\nu,\cdots]=\sum_{m=0}^\infty{d_{(m)}[\cdots]\over m!}\,\,
\dn{m}\left[{h\nu\over\kB T_0}\right]
\label{FokkerPlanck}
\end{equation}
where
\begin{equation}
\dn{m}[x]\equiv(-1)^m{d^m\nBB[x]\over d\ln[x]^m}\ .
\label{udistortions}
\end{equation}
By computing the moments of eq.~(\ref{DeltaExpansion}) one can express the
coefficients $d_{(m)}$ in terms of $g$, $\bar{T}$, and $u_{(n)}$ and vice versa
\begin{eqnarray}
{d_{(m)}\over1-g}&=&\sum_{n=0}^m{m!\over n!\,(m-n)!}\,
\ln\left[{\bar{T}\over T_0}\right]^{m-n}\,u_{(n)} \\ \nonumber
u_{(n)}&=&\sum_{m=0}^n{n!\over m!\,(n-m)!}\,
\ln\left[{T_0\over\bar{T}}\right]^{n-m}\,{d_{(m)}\over1-g}
\label{FokkerPlanckCoeffs}
\end{eqnarray}
(N.B. $u_{(0)}=1$, $u_{(1)}=0$).

\begin{figure}[tb]
\centerline{
\includegraphics[width=9cm]{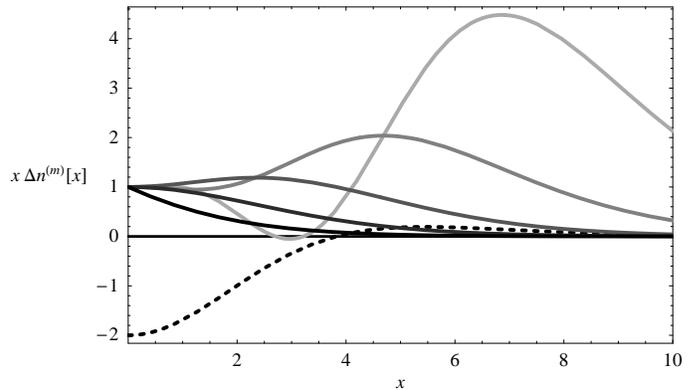}}
\caption{Shown are the spectral perturbation from the lowest order terms in the
Fokker-Planck expansion.  In particular the thick solid curves show
$x\,\dn{m}[x]$ defined in eq.~(\ref{udistortions}). As one goes from black to
gray the thick solid curves correspond to $m=0$ (a graybody distortion), $m=1$
(a temperature or Doppler distortion), $m=2$ (a $u$ distortion), $m=3$, and
$m=4$.  For reference a $y$-distortion $\dnSZ[x]=-3\dn{1}[x]+\dn{2}[x]$ is
shown as a dashed curve. Here $x=h\nu/(\kB T_0)$ where $T_0$ is a
temperature which will depend on the context.  Note that $x\,\Delta n$ is
proportional to the change in Rayleigh-Jeans temperature,
$T_{\rm RJ}\equiv(h\nu/\kB)\,n$.}
\label{ufig}
\end{figure}

Substituting eq.~(\ref{FokkerPlanckCoeffs}) into eq.~(\ref{FokkerPlanck}) and
using the Taylor series
\begin{equation}
\nBB\left[{h\nu\over\kB\,\bar{T}}\right]=
\sum_{j=0}^\infty {(-1)^j\over j!}\ln\left[{\bar{T}\over T_0}\right]^j
{\partial^j \over\partial\ln[\nu]^j}\nBB\left[{h\nu\over\kB\,T_0}\right]
\end{equation}
one finds an alternative Fokker Planck series
\begin{equation}
n[\nu,\cdots]=(1-g)\,\sum_{n=0}^\infty{u_{(n)}[\cdots]\over n!}\,\,
\dn{n}\left[{h\nu\over\kB\bar{T}}\right]\ .
\label{FokkerPlanckAlt}
\end{equation}
Eq.~(\ref{FokkerPlanckAlt}) involves the physically defined $\bar{T}$ rather
than the arbitrary $T_0$ in eq.~(\ref{FokkerPlanck}), the later is probably
more practically useful since precision spectral measurement are often with
respect to a reference blackbody.  Of course if $T_0=\bar{T}$, then
$d_{(m)}=(1-g)\,u_{(m)}$, the two equations are equivalent and $d_{(1)}=0$.

This formalism is most useful when the temperature distribution is narrow
enough that the first few terms of these series give an adequate approximation
to the spectrum.  A truncation of these series at $m\le N$ we call an order $N$
Fokker-Planck approximation.  Strictly speaking a Fokker-Planck approximation
corresponds to $N=2$.

\subsection{Fokker-Planck Asymptotes}
\label{sec:Spectrum:Asymptotes}

It is easy to compute the asymptotic form of the spectra. A low frequency
expansion is 
\begin{equation}
\dn{n}[x]=-{1\over2}+
\sum_{p=0}^\infty {B_{2p}\over(2p)!}\,(1-2p)^n x^{2p-1}
\end{equation}
where the $B_p$ are the Bernoulli numbers with values $B_0=1$, $B_1=-{1\over2}$,
while for integer $n\ge1$: $B_{2n+1}=0$ and
$B_{2n}={1\over2}-\sum_{k=0}^{n-1}{(2n)! B_{2k}\over(2k)!(2(n-k)+1)!}$.  A 
high-frequency expansion is
\begin{equation}
\dn{n}[x]=\sum_{r=1}^\infty p_n[r x]\,e^{-r\,x}\ .
\end{equation}
where the $p_n$ are order $n$ polynomials defined by
$p_n[x]=(-1)^n\,e^x {d^n\over d\ln[x]^n}e^{-x}$ with limiting values
$\lim_{x\rightarrow\infty}p_n[x]=x^n$ and $p_n[0]=\delta_{n,0}$.  Both
expansions have fairly good convergence properties.  From this one sees that
$\lim_{x\rightarrow0}\dn{m}[x]=1/x$ and
$\lim_{x\rightarrow\infty}\dn{m}[x]=x^m\,e^{-x}$.  The asymptotes are always
positive but $\dn{m}[x]$ will go negative for $n\ge4$.

\subsection{Power Law Moments}
\label{sec:Spectrum:PowerMoments}

The moments $\langle T^p\rangle$ are also useful, for example the bolometric
brightness is $\propto\langle T^4\rangle$ while the number flux of photons is
$\propto\langle T^3\rangle$.  Using 
eq.s~(\ref{DeltaExpansion},\ref{FokkerPlanckCoeffs}) one finds 
\begin{equation}
\langle T^p\rangle=T_0^p \sum_{m=0}^\infty {p^m\,d_{(m)}\over m!}
=(1-g)\,\bar{T}^p\sum_{n=0}^\infty {p^n\,u_{(n)}\over n!}\ .
\label{PowerMoments}
\end{equation}

\subsection{Sunyaev Zel'dovich Distortions}
\label{sec:Spectrum:SZ}

While the tSZ effect is not the focus of this paper, it does fit neatly into
the formalism just described.  The Kompane'ets equation which gives the
evolution of spectral distortions due to the tSZ effect, is an $N=2$
Fokker-Planck approximation to the collisional Boltzmann equation 
with Compton scattering by non-relativistic electrons. A cloud of hot electrons
illuminated by a blackbody spectrum with temperature $T_0$ will emit photons
with a distorted spectrum of the form
$n\approx\nBB\left[x\right]+y\,\dnSZ\left[x\right]$ where
\begin{eqnarray}
\label{ydistortion}
\dnSZ[x]&\equiv&{x\,e^x\over(e^x-1)^2}\,\left(x\,{e^x+1\over e^x-1}-4\right) \\
\nonumber
&=&-3\dn{1}[x]+\dn{2}[x]
\end{eqnarray}
which is called a $y$-distortion.  Here the ``$y$ parameter'' is given by a
line integral through the gas $y=\int d\tau \kB(\Te-T_0)/(\me c^2)$ where
$\tau=\int c\,\nel\,\sigmaT\,dt$ gives the Thomson optical depth,
$\sigmaT=6.65\times10^{-25}$cm$^2$ is the Thomson cross-section, $\nel$ is the
space density of free electrons, and $\kB\Te$ is 2/3 of the average kinetic
energy of the electrons (so $\Te$ gives the electron kinetic temperature for a
Maxwellian distribution).  The validity of eq.~(\ref{ydistortion}) is
$y,\tau\ll1$ and $\kB\Te,\,\kB\,T_0\,h\nu\ll\me c^2$. The latter requirements
give the condition for the width of the scattering kernel in frequency space to
be narrow, which is the requirement for a Fokker-Planck approximation.

One expects to find a varying $y$-distortion as one scans across the sky,
varying according to the density and temperature of hot gas along the
lines-of-sight. From eq.~(\ref{ydistortion}) one sees that
\begin{eqnarray}
\label{ymoments}
          g&=&1           +\calO[y^2]  \\ \nonumber
    \bar{T}&=&T_0\,(1-3\,y+\calO[y^2]) \\ \nonumber
  u=u_{(2)}&=&2\,y        +\calO[y^2]  \\ \nonumber
    u_{(n)}&=&0           +\calO[y^2]  \quad n\ge3 \ .
\end{eqnarray}
From eq.~(\ref{PowerMoments}) one recovers the well known result that the
change in the number flux of photons is unchanged:
${\Delta N\over N}=\langle T^3\rangle/T_0^3-1=0$, while the increase of
bolometric brightness is ${\Delta E\over E}=\langle T^4\rangle/T_0^4-1=4y$.

Various authors \cite{stebbins97a,stebbins97b,cl98,ikn98} have examined higher
order corrections in $\Thetae\equiv\kB\Te/(\me c^2)$, by expanding the
Boltzmann equation in powers of $\Thetae$, obtaining expressions like
eq.~(\ref{FokkerPlanck}) except that the coefficients are powers of $\Thetae$.
Note that this is only a rearrangement of
eq.s~(\ref{FokkerPlanck},\ref{FokkerPlanckAlt}) and the spectral shapes
obtained are linear combinations of the $\dn{m}$. For $\tau,\,h\nu\ll1$ and a
constant temperature gas, one express the spectral distortion by
eq.~(\ref{FokkerPlanckAlt}) where the $u_{(n)}$ are given by a power series in
$\Thetae$ beginning with power $\Thetae^{n-1}$.

The tSZ effect produces a generic Fokker-Planck distortion, however it can be
distinguished from other sources of spectral distortion, even those described
by a low order Fokker-Planck expansion.  The reason for this is that the tSZ
distortions are a one parameter ($\Thetae$) set of distortions, so that the
Fokker-Planck coefficients are not independent but are correlated.  From
eq.~(\ref{ydistortion}) one sees that for a pure $y$-distortion
($\calO[\Thetae^1]$), one expects $\ln[\bar{T}/T_0]=-3u_{(2)}/2$.  Even if it
were  contaminated by primordial anisotropies with
$u_{(2)}\ll\ln[\bar{T}/T_0]$, one could expect to find the unique tSZ
signature: 
\begin{equation}
{\rm Cov}\left[u_{(2)},\ln\left[{\bar{T}\over T_0}\right]\right]\approx
-{3\over2}\,{\rm Var}[u_{(2)}]
\label{ycorrelation}
\end{equation}
Other sources of distortion will not have this correlation.

\section{Boltzmann Equation}
\label{sec:Boltzmann}

So far we have just discussed mathematical descriptions of the spectral
distortions but not how they arise or their physical dynamics.  The equation
which describes the evolution of the distribution of photons in phase space
(space, time, direction, and frequency) in the presence of scattering is the
collisional  Boltzmann equation.  It can be written as a dynamical equation for
the occupation number $n$:
\begin{equation}
\label{Boltzmann}
{D\over Dt}n={d\tau\over dt} C
\end{equation}
where $D\over Dt$ is a convective derivative along a null geodesic, $C$ is a
dimensionless collision term, and ${d\tau\over dt}$ is the rate of increase of
scattering optical depth along the geodesic. This says that $n$ is conserved
along trajectories in phase space except for the effects of scattering (which
is really the definition of scattering).  In an arbitrary space-time the
trajectories are given by the geodesic equation, while $C$ and ${d\tau\over
dt}$ depends on the type of scattering.

The convective derivative may be written
\begin{equation}
\label{convective}
{D\over Dt}={\partial\over\partial t}
+{d\bfx\over dt}\cdot\nabla_{\bfx}
+{d\hatbfc\over dt}\cdot\nabla_{\hatbfc}
+{d\ln[\nu]\over dt}{\partial\over\partial\ln[\nu]}\ .
\end{equation}
This equation is purposely written in 3+1, Galilean language, but is applicable
in an arbitrary space-time when one chooses a global rest-frame (see
appendix~\ref{app:Frames} for the specifics).  Here ${d\bfx\over dt}$ give the
flow in position space, while ${d\hatbfc\over dt}$ gives the flow in direction
space, \ie the bending of light.  These two terms are generally what is known
as ``lensing'', but in this paper the main interest is in the 4th term, \ie the
flow in frequency space.

\subsection{Redshift}
\label{sec:Boltzmann:redshift}

In appendix~\ref{app:Frames} it is shown that
\begin{equation}
{d\ln[\nu]\over dt}={d\ln[1+z]\over dt}=
-\hatbfc\cdot(\grad\bfv)\cdot\hatbfc+\hatbfc\cdot\bfa
\label{redshift}
\end{equation}
where $\grad\bfv$ and $\bfa$ are respectively the spatial velocity gradient and
proper acceleration of the global frame we are using, while $z$ is the {\it
redshift} ($1+z$ is only defined up to a multiplicative constant).

This is not an expression for redshift which may be familiar to you, but this
non-perturbative expression is completely intuitive from well-known,
perturbative, Newtonian and cosmological results. The 1st term on the
right-hand-side (rhs) gives the line-of-sight component to the velocity
gradient of the observer, so this terms can be interpreted as a Doppler
redshift having to do with differences in the observers' velocities.  One can
interpret $-\bfa$ as the gravitational acceleration ($\bfg=-\bfa$ in Newtonian
gravity) and the $-\hatbfc\cdot\bfa$ term gives the gravitational redshift as
the photons move against or with the ``force of gravity''.  In different frames
the amount of Doppler and gravitational redshift will differ, but if one
considers two fixed endpoints the gravitational redshift plus the correction
for the Lorentz boosts between different frames will always agree.

	Eq.~(\ref{redshift}) also produces a non-perturbative expression for
``cosmological'' space-times.  Consider the metric 
\begin{equation}
g_{\alpha\beta}=
{\diag[-e^{2\phi},e^{-2\psi},e^{-2\psi},e^{-2\psi}]\over (1+\zc[\eta])^2}
\label{frw}
\end{equation}
where $\zc[\eta]$ is the cosmological redshift, $\eta$ is the conformal time
coordinate, $\bfx$ are the spatial coordinates, while $\phi[\bfx,\eta]$ and
$\psi[\bfx,\eta]$ give arbitrarily large perturbations from a flat 
Friedman-Robertson-Walker (FRW) space-time (if $\phi=\psi=0$ this would be a
flat FRW cosmology). In the coordinate frame, the solution of
eq.~(\ref{redshift}) is
\begin{equation}
{1+z_2\over1+z_1}
={1+\zc[\eta_2]\over1+\zc[\eta_1]}\,
e^{\phi_1-\phi_2+\int_{\eta_1}^{\eta_2} d\eta\,(\dot\phi+\dot\psi)}
\end{equation}
for the redshift between two points 1 and 2 along any null geodesic,
$\bfr[\eta]$, in this space-time.  Here $\dot{}={\partial\over\partial\eta}$ and
$\phi_i=\phi[\bfr[\eta_i],\eta_i]$.  Note that the logarithm in
eq.~(\ref{redshift}) means that sum of terms on the rhs translate into a
product of terms in the expression for $1+z$.  Note that if $\phi=\psi=0$ then
then the coordinate frame is free-falling, so $\bfa=0$, and there is only a
Doppler redshift.  More generally the Doppler term is divided into two factors:
${1+\zc[\eta_2]\over1+\zc[\eta_1]}$ and $e^{\int d\eta\,\dot{\psi}}$; the first
is known as the {\it cosmological redshift} given by the velocity gradient when
$\phi=\psi=0$ and the 2nd term arises because when $\dot{\psi}\ne0$ the
coordinates expand or contract leading to additional Doppler shifts for
observers moving with the coordinates.  The gravitational redshift is also
divided into two factors, by splitting $\hatbfc\cdot\bfa$ into a term which a
perfect derivative along the null geodesic and yields $e^{\phi_1-\phi_2}$ and
the remainder which yields $e^{\int d\eta\,\dot{\phi}}$.  Mimicing perturbative
cosmology terminology one would call $e^{\phi_1-\phi_2}$ the {\it Sachs-Wolfe}
(SW) redshift and $e^{int_{\eta_1}^{\eta_2} d\eta\,(\dot\phi+\dot\psi)}$ the
{\it integrated Sachs-Wolfe} (ISW) redshift. These expressions are remarkably
similar to the well-known result for linear perturbations; even though this is
completely non-perturbative.  Of course eq.~(\ref{frw}) does not include the
most general perturbations from a flat FRW space-time (\ie no vector or tensor
perturbations), and one still has to solve for the trajectory $\bfr[\eta]$
first.

\subsection{Thomson Scattering in the Baryon Frame}
\label{sec:Boltzmann:Thomson}

Now since we are using a particular global frame in which to write the
Boltzmann equation, it is most convenient to use the baryon frame in which the
electrons are at rest.  An important point here is that we are considering only
the spectral distortion for a cold gas of electrons, and specifically ignoring
the electron velocity dispersion, which will also produce a spectral distortion
via the tSZ effect.  In some, but not all cases, the tSZ effect will dominate
the spectral distortion considered here, but as the anisotropy-scattering
coupling considered here has usually been neglected it is worthwhile to
consider it on it's own.

Here we consider not only cold electrons but also the limit of low energy
photons \ie $h\nu\ll\me c^2$.  In this case the cross-section for scattering an
unpolarized beam is proportional to
\begin{equation}
\label{UnpolarizedCrossSection}
\Sigma[\hatbfc,\hatbfc']={3\over16\pi}\left(1+(\hatbfc\cdot\hatbfc')^2\right)
\end{equation}
Using the the notation
\begin{equation}
\label{SigmaOperator}
\SigmaOp F[a,b']\equiv
\int d^2\hatbfc'\,\Sigma[\hatbfc,\hatbfc']\,F[a[\hatbfc],b[\hatbfc']]\ .
\end{equation}
the Boltzmann eq.~(\ref{Boltzmann}) becomes
\begin{equation}
\label{nBoltzmann}
{D\over Dt}n={d\tau\over dt}\,\SigmaOp (n'-n) .
\end{equation}
Henceforth a $'$'d quantity indicates that it is evaluated at direction
$\hatbfc'$.  Note that $\SigmaOp$ is a linear convolution operator since
$\SigmaOp1=1$.

\subsection{Temperature Transform of Boltzmann Equation}
\label{sec:Boltzmann:Temperature}

Substituting eq.~(\ref{temperaturetransform}) into eq.~(\ref{nBoltzmann}) one
finds that the Boltzmann equation for $q[\ln[T],\hatbfc,\bfx,t]$ in the baryon
frame is
\begin{equation}
{Dq\over Dt}={d\tau\over dt}\,\SigmaOp\,(q'-q)
\label{TemperatureBoltzmann}
\end{equation}
where the convective derivative in temperature space is
\begin{equation}
{D\over Dt}={\partial\over\partial t}+{d\hatbfc\over dt}\cdot\nabla_\bfx
+{d\hatbfc\over dt}\cdot\nabla_\hatbfc
+{d\ln[1+z]\over dt}{\partial\over\partial\ln[T]}\ .
\label{TemperatureConvection}
\end{equation}

\subsection{Boltzmann Equation of Moments}
\label{sec:Boltzmann:Moments}

When one deal with temperature moments the temperature dependence is already
``removed'' so in what follows we use the convective derivative
\begin{equation}
{\calD\over\calD t}={\partial\over\partial t}+{d\bfx\over dt}\cdot\nabla_\bfx
+{d\hatbfc\over dt}\cdot\nabla_\hatbfc
\end{equation}
and any effect of redshifting is shifted to the rhs of the equation.  Since $g$
and $u_{(n)}$ are Lorentz invariant there will be no redshifting terms for
these quantities. With this convention if we substitute eq.~(\ref{logmoments})
into eq.~(\ref{TemperatureBoltzmann}) then we find
\begin{equation}
{\calD \eta_{(n)}\over\calD t}=n\,{d\ln[1+z]\over dt}\,\eta_{(n-1)}
+{d\tau\over dt}\,\SigmaOp(\eta_{(n)}'-\eta_{(n)})\ .
\label{MomentBoltzmann}
\end{equation}
This equation tells us that the $\eta_{(n)}$ moments depend only on moments
with smaller $n$ so that one may without loss of accuracy truncate the
evolution of the moments at any order $n$.

What one really wants is the evolution of $g$, $\bar{T}$, the central moments
$u_{(n)}$; which one obtains by combining
eq.s~(\ref{grayness},\ref{meantemp},\ref{centralmoments}) with
eq.~(\ref{MomentBoltzmann}) to obtain
\begin{eqnarray} \nonumber
\label{CentralBoltzmann}
{\calD g\over\calD t}&=&{d\tau\over dt}\,\SigmaOp(g'-g) \\ \nonumber
{\calD\ln[{\bar{T}\over1+z}]\over\calD t}&=&
{d\tau\over dt}\,\SigmaOp\left(
{1-g'\over1-g}\,\ln\left[{\bar{T}'\over\bar{T}}\right]\right) \\ \nonumber
{\calD u_{(n)}\over\calD t}&=&{d\tau\over dt}\,
\SigmaOp\left({1-g'\over1-g}\,U_{(n)}\right) \\ \nonumber
U_{(n)}&=&\left(\sum_{m=0}^n{n!\over m!(n-m)!}\,
      \ln\left[{\bar{T}'\over\bar{T}}\right]^{n-m}\,u_{(m)}'\right) \\
&& -n\,\ln\left[{\bar{T}'\over\bar{T}}\right]\,u_{(n-1)}-u_{(n)}
\end{eqnarray}
Eq.s~(\ref{CentralBoltzmann}), along with the polarized version,
eq.s~(\ref{TensorBoltzmannCentral}), are the main results of this paper.

As with the $\eta_{(n)}$ this set of coupled equations can be truncated at any
order $N$, leading to an accurate representation of all the spectral components
$\dn{n}$ up to order $N$.  Of course a truncation does not give a complete
description of the spectrum.  Note that this is not a perturbation expansion,
rather these are completely non-perturbative equations for arbitrary
space-times and for large spectral distortions.  The assumptions are cold
electrons $\kB\Te\ll\me c^2$, soft photons $h\nu\ll\me c^2$, only Thomson
scattering, and unpolarized light.  All of these assumptions are liable to be
real limitations in applicability.  The lack of polarization in these equations
was done for simplicity, the formulae including polarization is given in
appendix~\ref{app:Polarization}.  In most applications polarization will lead
only to a small correction and in some cases polarization is completely
negligible.  These equations are tied to the gas frame which may not be most
convenient for every application.  It is relatively simple to translate these
equations into a different frame: note that $g$ and $u_{(n)}$ are frame
invariant while $t$, $\hatbfc$, $\bar{T}$ and all the dot products are not.

\subsubsection{Boltzmann Equation for $u$}
\label{sec:Boltzmann:Moments:u}

As we shall see the most important distortions are the lowest order ones, \ie
$g$, $\bar{T}$ and $u=u_{(2)}$.  The equations for $g$ and $u$ are explicit in
eq.s~(\ref{CentralBoltzmann}) and here is the explicit equation for
$u=u_{(2)}$:
\begin{equation}
{\calD u\over\calD t}={d\tau\over dt}\,\SigmaOp\left({1-g'\over1-g}\,
\left((u'-u)+\ln\left[{\bar{T}'\over\bar{T}}\right]^2\right)\right)\ .
\end{equation}
One sees that apart from convection, the $u$ distortion is sourced by spectral
anisotropy, $u'-u$, and by temperature anisotropy $\ln[\bar{T}'/T]$.  This is
illustrative of all the moments moments $u_{(n)}$ in that
$\ln[\bar{T}'/\bar{T}]$ can directly produce $u_{(n)}$ and in that the equation
is linear in $u_{(n)}$.  For $n>2$ the scattering term is more complicated and
always includes non-linear coupling between $\ln[\bar{T}'/\bar{T}]$ and
$u_{(m)}$.

\section{Spectral Dynamics}
\label{sec:SpectralDynamics}

\subsection{Temperature and Spectral Anisotropies}
\label{sec:SpectralDynamics:Anisotropies}

Here the word {\it anisotropy} is used to mean any function of the
$n[\nu,\hatbfc,\cdots]$ at which is zero when $n$ is independent of $\hatbfc$.
This includes functions of $q$, $\eta_{(n)}$, $g$, $\bar{T}$, or $u_{(m)}$
which are zero when there is no $\hatbfc$ dependence.  It is clear for
eq.s~(\ref{TemperatureBoltzmann},\ref{MomentBoltzmann}) that the collision term
in the Boltzmann equation is an anisotropy.  One can also see this most clearly
for eq.~(\ref{CentralBoltzmann}) by grouping the $u_{(m)}$ dependence of
$U_{(n)}$, \ie decomposing $U_{(n)}=\sum_{m=0}^n U_{(n,m)}$ where
\begin{eqnarray}
U_{(n,0)}  \hskip-8pt&=&\hskip-8pt \ln\left[{\bar{T}'\over\bar{T}}\right]^n 
                                                       \\ \nonumber
U_{(n,1)}  \hskip-8pt&=&\hskip-8pt 0                   \\ \nonumber
U_{(n,n)}  \hskip-8pt&=&\hskip-8pt u_{(n)}'-u_{(n)}    \\ \nonumber
U_{(n,n-1)}\hskip-8pt&=&\hskip-8pt 
       n\,\ln\left[{\bar{T}'\over\bar{T}}\right]^{n-m}\,(u_{(n-1)}'-u_{(n-1)})
             \hskip11pt n\ge3                          \\ \nonumber
\tilde{U}_{(m,n)}\hskip-8pt&=&\hskip-8pt
{n!\over m!\,(n-m)!}\ln\left[{\bar{T}'\over\bar{T}}\right]^{n-m} u_{(m)}'
\hskip 5pt {n\ge3 \atop m\in[2,n-2]}\ .
\label{Udecompose}
\end{eqnarray}
In eq.s~(\ref{CentralBoltzmann},\ref{Udecompose}) contains terms like
$\ln\left[\bar{T}'\over\bar{T}\right]$, which is a {\it temperature
anisotropies} and terms like $g-g'$, and $u_{(n)}'-u_{(n)}$ which are
{\it spectral anisotropies}.  Note again that if the spectral and temperature
anisotropies are zero then the scattering term is zero.

\subsection{Initial Conditions and Origin of Spectral Distortions}
\label{sec:SpectralDynamics:Origin}

At very early times electron and atomic scattering is sufficient to 
nearly completely isotropize and thermalize the photon distribution to an
isotropic blackbody, so the initial conditions (ICs) are
\begin{equation}
\label{ICs}
\lim_{t\rightarrow0}\{g,\,\ln\left[{\bar{T}'\over\bar{T}}\right],\,u_{(n>0)}\}
=0 \ .
\end{equation}
So the spectral distortions evolve from zero and the reason they exist is
because of inhomogeneities.  Starting with these ICs one sees from
eq.~(\ref{CentralBoltzmann}) that
\begin{equation}
\label{zerog}
g=0\ .
\end{equation}
Note however that this result does not take into account other radiative
foreground which can cause $g$ to vary from zero.

The initial growth of temperature anisotropy is given by
\begin{eqnarray}
\label{anisotropyIC} \nonumber
\lim_{t\rightarrow0}{\partial
    \ln\left[{\bar{T}[\hatbfc_2,\cdots]\over\bar{T}[\hatbfc_1,\cdots]}\right]
    \over\partial t}
&=&{d\ln\left[{1+z[\hatbfc_2,\cdots]\over1+z[\hatbfc_1,\cdots]}\right]\over dt}
\\ \cr
&&\hskip-125pt=(\hatbfc_2-\hatbfc_1)\cdot\bfa_\rmp
+\hatbfc_1\cdot(\nabla\bfv)\cdot\hatbfc_1
-\hatbfc_2\cdot(\nabla\bfv)\cdot\hatbfc_2\ .
\end{eqnarray}
Thus anisotropies in the baryon frame are initially caused by acceleration of
the gas ($\bfa_\rmp\ne0$ usually due to pressure gradients) and/or by
anisotropic gas velocity gradients (shear). Shear will always be a consequence
of inhomogeneities in the universe and will inevitably lead to temperature
anisotropies.  One sees from eq.s~(\ref{CentralBoltzmann}) that anisotropy will
lead to time varying, and hence nonzero, spectral distortions $u_{(n)}$.  Note
however that scattering tends to damp existing temperature anisotropies toward
zero so the amount of anisotropy and associated spectral distortions are highly
suppressed until scattering turns off at recombination.

\subsection{Perturbative Analysis}
\label{sec:SpectralDynamics:Perturb}

Consider a one parameter family of solutions to the full equations-of-motion
(EoMs) for matter and gravity as well as the initial conditions (ICs).  When
$\epsilon=0$ the EoMs and ICs give the unperturbed background cosmology, and
more generally $\epsilon$ gives the amplitude of the perturbation from the
background solution.  One can Taylor series the EoMs and ICs about $\epsilon=0$
and then solve them at each order using the lower order solutions.  The ICs may
be stochastic.  One can Taylor series in $\epsilon$ about 0 any quantity which
depends on the solution, \eg $Q=\sum_{p=0}^\infty Q^{\{i\}}\epsilon^p$ If the
smallest $p$ for which the $Q^{\{n\}}\ne0$ is $N$ then one says that $Q$ is a
{\it perturbation variable of order $N$} and denote this by $Q\sim\calO[N]$.
For two quantities $P$ and $Q$, if $P-Q\sim\calO[N]$ then one sees that
$P^{\{M\}}=Q^{\{M\}}$ for all $M<N$.

The solution of our EoMs, eq.~(\ref{zerog}), tells us that $g\sim\calO[\infty]$
although, as mentioned above, additional radiative processes will cause a
non-zero $g$ at some order.  Since an unperturbed cosmology is by definition
homogeneous and everywhere isotropic one must have
\begin{equation}
\ln\left[{\bar{T}'\over\bar{T}}\right]^{\{0\}}
=(\nabla_\bfx n)^{\{0\}}=(\nabla_\hatbfc n)^{\{0\}}
={d\hatbfc\over dt}^{\{0\}}=0 \ .
\end{equation}
We know from \S\ref{sec:SpectralDynamics:Origin} that
$\ln\left[{1+z[\hatbfc_2,\cdots]\over1+z[\hatbfc_1,\cdots]}\right]\sim\calO[1]$
and hence $\ln[\bar{T}'/\bar{T}]\sim\calO[1]$.  It follows that
$\ln[\bar{T}'/\bar{T}]^n\sim\calO[n]$ and given the ICs of eq.~(\ref{ICs}) one
can see from eq.s~(\ref{CentralBoltzmann}) that $u_{(n)}\sim\calO[n]$ for
$n\ge2$ (by definition $u_{(0)}=1\sim\calO[0]$ and
$u_{(1)}=0\sim\calO[\infty]$).  Thus for perturbation theory at a given order
$N$ one need only consider $u_{(n)}$ for $n\le N$.  In linear theory, $N=1$,
there is no spectral distortion, only temperature anisotropy. From the
scattering of anisotropies the lowest order spectral distortion is
$u=u_{(2)}\sim\calO[2]$, \ie spectral distortions only appear in 2nd order
perturbation theory.

These conclusions depend on our EoMs which ignore certain radiative processes.
In particular only non-relativistic Thomson scattering with $h\nu\, \kB
T_\rme\ll m_\rme\,c^2$ has been assumed.  Finite temperature and frequency
corrections are small, but so are the spectral distortions.  Long before
recombination these conditions are violated but the scattering tends to damp
spectral distortions and I estimate these corrections are relatively small
although formally they do decrease the order of the spectral distortions.  Even
at $\calO[0]$ there will be finite temperature differences between the
electrons and photons which will lead to a 0th order isotropic spectral
distortion from the tSZ effect, \ie the isotropic part of $\langle
u_{(n)}\rangle_{\hatbfc}\sim\calO[0]$, an effect not included in our
EoMs. However the amplitude of these terms does decrease rapidly with $n$
because it is non-relativistic tSZ.  Furthermore even $\langle
u_{(2)}\rangle_{\hatbfc}$ is very small.  Spectral anisotropies from this
effect only arise through coupling to temperature anisotropies, so the formally
anisotropic part, is of higher order $u_{(n)}-\langle
u_{(n)}\rangle_{\hatbfc}\sim\calO[1]$, although this again is a small effect
and decreases rapidly with $n$.

The most important spectral distortion which has been ignored, is from tSZ at
low $z$ caused by shock-heated gas.  Shock heating is, arguably, a
non-perturbative process, but nevertheless may lead to the largest spectral
distortion.

\subsection{${\Delta T\over T}$}
\label{sec:SpectralDynamics:DeltaToverT}

One normally denotes a temperature anisotropy by $\Delta T/T$.  What this
specifically means may vary, but in many cases
${\Delta T\over T}=(\Tb-T_0)/T_0$ where $T_0$ is some reference temperature and
$\Tb$ is the brightness temperature at a particular frequency. If $g=u_{(n)}=0$
but $\bar{T}\ne0$ then from eq.s~(\ref{FokkerPlanck}-\ref{FokkerPlanckCoeffs})
the distorted spectrum is
\begin{equation}
n={1\over e^x-1}\,
\left(1+\ln\left[{\bar{T}\over T_0}\right]\,{x\,e^x\over e^x-1}\right)
\qquad x={h\nu\over\kB T_0}\ .
\end{equation}
If one Taylor expands the dependence of $\Tb$ on 
$\ln[\bar{T}/T_0]$ about 0 one finds
\begin{eqnarray}
{\Delta T\over T}&\equiv&{\Tb-T_0\over T_0}=\ln\left[{\bar{T}\over T_0}\right] 
\\ \nonumber && \hskip-30pt
+\left(1-{x\over2}+{x\over e^x-1}\right)\,\ln\left[{\bar{T}\over T_0}\right]^2
+\calO\left[\ln\left[{\bar{T}\over T_0}\right]^3\right]\ .
\end{eqnarray}
Of course $\bar{T}\sim\calO[0]$ and by the assumed symmetry
$\bar{T}^{\{0\}}=\Tc[t]$ which is independent of $\bfx$ and $\hatbfc$.  It then
follows that $(\Delta T/T)^{\{1\}}=\ln[\bar{T}/\Tc[t]]^{\{1\}}$ and
\begin{equation}
\ln\left[{\bar{T}'\over\bar{T}}\right]^{\{1\}}
={{\Delta T\over T}'}^{\{1\}}-{\Delta T\over T}^{\{1\}}\ .
\end{equation}
\ie for small anisotropies $\ln[\bar{T}'/\bar{T}]$ reduces to the conventional
definition of temperature anisotropy!

\subsection{Lowest Order Spectral Distortions}
\label{sec:SpectralDynamics:Lowest}

Given homogeneity and isotropy of the background solution and using
${d\ln[1+z]\over dt}^{\{0\}}=-H[t]$ where $H$ is the Hubble parameter, the only
non-trivial part of 0th order Boltzmann equations is
\begin{equation}
{d\ln[\Tc[t]]\over dt}=-H[t]\ .
\end{equation}
Defining the cosmological redshift 
$\zc[t]=-1+e^{\int_t^{t_0}\,dt'\,H[t']}$, the solution is the usual
temperature-redshift relation $\Tc[t]=\Tc[t_0]/(1+\zc[t])$. 

The only non-trivial 1st order equation is the usual linearized Boltzmann
equation for ${\Delta T\over T}$ in a frame comoving with the baryons.  This is
what is usually used to calculate CMBR anisotropies.  What is new here only
enters at $\sim\calO[2]$, which is where the lowest order spectral distortion
arise.  These lowest order distortion are solutions to the 2nd order equations
for $u=u_{(2)}$:
\begin{eqnarray}
\label{uEquation}
&&\hskip-20pt{\partial u^{\{2\}}\over\partial t}
+\left({d\bfx\over dt}\cdot\nabla_\bfx\right)^{\{0\}}u^{\{2\}}
+\left({d\hatbfc\over dt}\cdot\nabla_\hatbfc\right)^{\{0\}}u^{\{2\}}
\\ \nonumber &&\hskip-20pt
={d\tau\over dt}^{\{0\}}\,\SigmaOp\left(u'^{\{2\}}-u^{\{2\}}+
\left({{\Delta T\over T}^{\{1\}}}'-{\Delta T\over T}^{\{1\}}\right)^2\right)
\end{eqnarray}
The 0th order gradient operators depends on the coordinate system one uses in
the background cosmology, but if one is perturbing from a flat, $\Omega_0=1$,
universe then one can choose comoving Euclidean coordinates normalized to
$\zc=0$ so that $d\hatbfc/dt^{\{0\}}=0$ and
$\left((d\bfx/dt)\cdot\nabla_\bfx\right)^{\{0\}}
=c\,(1+\zc[t])\,\hatbfc\cdot\nabla_{\rm co}$.  Eq.~(\ref{uEquation}) is derived
in the frame comoving with the baryons.  In 1st order perturbation theory, to
transform to another frame only involves correcting the dipole component of
temperature anisotropy for the Doppler shift caused by the velocity of the
baryons in the new frame, since angles are not changed to 1st order and $u$ is
frame independent.

Thus in an arbitrary frame (\ie gauge) to 1st order make the substitution
\begin{equation}
\label{DopperShift}
 {\Delta T\over T}^{\{1\}}[\hatbfc,\bfx,t]
={\Delta T\over T}^{\{1\}}[\hatbfc,\bfx,t]
\biggl|_{\rm baryon\atop frame}-{\hatbfc\cdot\bfvb^{\{1\}\over c}}
\end{equation}
where $\bfvb$ is the baryon velocity in the new frame.  The lowest order
equation for $u$ in the new frame is thus
\begin{eqnarray}
\label{uEquationSimple}
&&\hskip-20pt{\partial u\over\partial t}
+c\,(1+\zc)\,\hatbfc\cdot\nabla_{\rm co}\,u
\\ \nonumber &&\hskip-20pt
={d\overline{\tau}\over dt}\SigmaOp\left(u'-u+
\left({\Delta T\over T}'-{\Delta T\over T}
      +{(\hatbfc-\hatbfc')\cdot\bfvb\over c^2}\right)^2\right)
\end{eqnarray}
where the order superscripts have been dropped, and $\overline{\tau}$ is used
for the 0th order optical depth, to emphasize that it is spatially constant.
This equation is sufficient for most applications, since the anisotropies are
small, the spectral distortions are even smaller, and the lowest order spectral
distortion will be, by far, the largest.

\subsection{Single Scattering}
\label{sec:SpectralDynamics:Single}

A commonly used approximation where the optical depth to scattering is small,
for example in the modeling of tSZ effect for clusters of galaxies, is to
assume the photons undergo at most one scattering, and that the light incident
on gas is isotropic, unpolarized, and spectrally undistorted.  Here allow the
incident light to have anisotropic $\bar{T}$ but zero $g$ and $u_{(n)}$ (for
$n\ge1$).  Under these assumption most of the terms in
eq.s~(\ref{CentralBoltzmann}) are zero and one can express the solution as an
integral along the photon trajectory:
\begin{equation} \nonumber
\label{LowTau}
u_{(n)}=\int d\tau\,\SigmaOp\ln\left[{\bar{T}'\over\bar{T}}\right]^n\ .
\end{equation}
The polarized version of this integral comes by substituting $g=0$ and
$\Ut{n}=\delta_{n,0}\It/2$ into
eq.s~(\ref{TensorBoltzmannCentral},\ref{Atensor}) and integrating to
obtain 
\begin{equation}
\Ut{n}={3\over4}\int d\tau\,
\AngleAverage{\ln\left[{\bar{T}'\over\bar{T}}\right]^n
\It\hskip-2pt\cdot\hskip-2pt\It'\hskip-2pt\cdot\hskip-2pt\It}\ .
\end{equation}
The trace of this is eq.~(\ref{LowTau}).  Both of these integrals can give a
good approximation to all of the moments in some situations.

\subsection{Adding Thermal SZ}
\label{sec:SpectralDynamics:tSZ}

Much of the formalism developed in this paper is appropriate for arbitrarily
large spectra distortions but restricted to cold electrons.  The tSZ effect
described in \S\ref{sec:Spectrum:SZ} will provide a correction to the collision
term of the Boltzmann equation due to finite velocity dispersion of the
electrons.  This effect is most well known for the case of small spectral
distortions and non-relativistic electron velocity
dispersions\footnote{Strictly speaking the tSZ effect is a relativistic effect
since it scales like $(v/c)^2$ of the electrons.  By non-relativistic tSZ
effect one means that $v\ll c$ so the tSZ effect is small.} in which case one
gets a $y$-distortion.  In this small distortion limit one can simply add the
$y$ distortion as a collision term in addition to the Thomson scattering term
already included.  To do this one need only note the the relation of $y$ to the
central moments given in eq.~(\ref{ymoments}).  We see that there is no
modification needed for the Boltzmann equation of $g$ and $u_{(n)}$ for
$n\ge3$.  The only modifications are for $\bar{T}$ and $u=u_{(2)}$, which in
the baryon frame are then corrected by
\begin{eqnarray}
\label{addy}
{\calD\ln[{\bar{T}\over1+z}]\over\calD t}&=&\cdots-3\,{dy\over dt}\\ \nonumber
{\calD u\over\calD t}&=&\cdots+{dy\over dt}
\end{eqnarray}
The validity of this equation is only for small distortions so one should
restrict oneself to lowest order terms as in eq.~(\ref{uEquationSimple}).  The
usual expression for the total $y$ along a trajectory is $y=\int
dt\,(d\tau/dt)\,\kB\,(\Te-T_\gamma)/(\me c^2)$ where $\tau$ is the Thomson
optical depth used previously, $\Te$ is the electron temperature, and
$T_\gamma$ is the photon temperature assumed nearly thermal and isotropic.
More generally, when the electrons are non-relativistic but isotropic, one can
use $\kB\Te={1\over3}\me\overline{\ve^2}$, where $\overline{v_\rme^2}$ is the
velocity dispersion of the electrons.  For $T_\gamma$ one wants the average
temperature of the photons scattered into the beam so one should use
$T_\gamma=\SigmaOp\bar{T}'$.  Thus one should use
\begin{equation}
{dy\over dt}
={d\tau\over dt}\,{\kB\,\SigmaOp(\Te-\bar{T}')\over\me c^2}\ .
\end{equation}

One should expect similarities between the tSZ effect and the scattering of
anisotropies.  This is because underlying both is non-relativistic Compton
scattering \ie Thomson scattering.  The relativistic corrections to the Thomson
cross-section in the center-of-mass frame only enters when the center-of-mass
kinetic energy $KE_{\rm cm}$ approaches $\me c^2$.  For microwave photons this
would require electron energies of several TeV.  For lower energies the tSZ
effect is just the sum of Thomson scatterings with varying Lorentz boosts
depending on the motion of the electrons.  Since $g$ and $u_{(n)}$ are Lorentz
invariant, to compute even the relativistic tSZ effect one only needs to adjust
the formulae for a sum of Lorentz boosts.  One trivial result is that $g=0$ is
also fixed point for tSZ effect.  This result does not depend on the thermality
of the electron velocity distribution, and holds even when the electrons are
relativistic.

\section{Average $u$}
\label{sec:uAverage}

Unlike for anisotropies which have, by definition, zero mean, the quantity $u$
must be positive.  It is therefore interesting to derive an expression for the
spatial average of $u$, at a given $t$ \ie $\overline{u}[t]\equiv\langle
u\rangle_{\bfx,\hatbfc}$. For the lowest order dynamics this is straightforward
and it is easiest to do so in terms of the angular power spectrum of
anisotropies $\hat{C}_l[\bfx,t]$ at each space-time point defined in the usual
way:
\begin{eqnarray} \nonumber
a_{(l,m)}[\bfx,t]&\equiv&\int d^2\hatbfc\,
Y_{(l,m)}[\hatbfc]^*{\Delta T\over T}[\hatbfc,\bfx,t] \\
\hat{C}_l[\bfx,t]&\equiv&{1\over2l+1}\sum_{m=-l}^l|a_{(l,m)}[\bfx,t]|^2\ .
\end{eqnarray}
Here the $Y_{(l,m)}[\hatbfc]$ are spherical harmonic functions.  Taking the
average of eq.~(\ref{uEquationSimple}), including the tSZ correction of
eq.~(\ref{addy}), and integrating with initial condition $\overline{u}[0]=0$
one finds
\begin{eqnarray}
\label{uAverage}
\overline{u}[t]&=&\ySZ[t]+\uv[t]+\udT[t] \\ \nonumber
\ySZ[t]&=&\int_0^t dt\,{d\overline{\tau}\over dt}
{\kB\langle\Te-\bar{T}\rangle_{\hatbfc,\bfx}\over\me c^2}    \\ \nonumber
\uv[t]&=&\int_0^t dt\,{d\overline{\tau}\over dt}
{2\over3}{\langle|\bfv_{\gamma\rm b}|^2\rangle_\bfx\over c^2}\\ \nonumber
\udT[t]&=&\int_0^t dt\,{d\overline{\tau}\over dt}
\sum_{l=2}^\infty{2l+1\over2\pi}\,\left(1-{1\over10}\delta_{l,2}\right)
\langle \hat{C}_l\rangle_\bfx
\end{eqnarray}
Here $\bfv_{\gamma\rm b}[\bfx,t]$ is the velocity of the baryons in the CMB
frame.  Here the scattering of the dipole anisotropy in $\uv$ has been split
from the scattering of the higher $l$ harmonics in $\udT$.  In the baryon frame
${2\over3}|\bfv_{\gamma\rm b}|^2/c^2=(2 1+1)\hat{C}_1/(2\pi)$ so $\uv+\udT$ is
the expected mean square anisotropy in the baryon frame, not including the
monopole ($l=0$), including the dipole ($l=1$), and subtracting ${1\over10}$ of
the quadrapole ($l=2$).  One expects that the random nature of cosmological
inhomogeneities is ergodic so one can replace the spatial averages with an
average over realizations, and use $C_l$'s computed by software like CMBFAST
\cite{CMBFAST}.

The average $u$ is effected by both scattering of anisotropies and the tSZ
effect, in contrast since
\begin{equation}
\AngleAverage{\SigmaOp\ln\left[{\bar{T}'\over\bar{T}}\right]}=0
\end{equation}
one finds
\begin{equation}
\label{Tave}
\langle\bar{T}\rangle_{\hatbfc,\bfx}[t]
\propto(1+\zc[t])\,e^{-3\,\ySZ[t]}
\end{equation}
which is only effected by the tSZ effect and not scattering of anisotropies.

\subsection{Degeneracy of $\uv$ and $\ySZ$}

These mean spectral distortion for the scattering of anisotropies, $\uv+\udT$,
cannot be disentangled from the tSZ $y$-distortion because we have no {\it a
priori} knowledge of what the pre-tSZ $\bar{T}$ was \ie we cannot use
eq.~(\ref{Tave}) to determine $\ySZ$ and then subtract it from $u$.  One way to
break this degeneracy is to instead consider anisotropies in $u$ and $\bar{T}$
to see what the covariance in these two quantities are.  For a pure tSZ one
expects the relation eq.~(\ref{ycorrelation}), but the scattering effect will
decrease the covariance since scattering does not correlate $\bar{T}$ with $u$.
The primary CMB temperature anisotropies will be a significant source of
contamination and one might find that there is not enough sky to provide good
enough statistics to measure the covariance accurately enough.

\subsection{$\bar{u}$ from Early Times}
\label{earlyu}

At times around and before recombination, $\zc\simgt1100$, the anisotropies are
small $\simlt 10^{-5}$ but the optical depth is very high and the electron and
photon temperature is nearly in equilibrium so the contribution of these epochs
to eq.~(\ref{uAverage}) is not obvious.  However I expect that the late-time
rather that early-time contribution to $\uv$ will dominate.

\subsection{$\bar{u}$ from Late Times}
\label{lateu}

Soon after recombination the optical depth becomes very small until the time of
reionization, believed to be $\zrei\sim 10$; integrating to a total optical
depth of $\sim0.07$.  In standard cosmologies, after reionization by far the
largest contribution to the anisotropies is the dipole from the relative
velocity of the baryons and the photons, \ie the $v_{\gamma\rm b}$ term.  A
large tSZ contribution to $\uv$ is also produced at late times as the gas will
be adiabatically and shock heated as non-linear collapse of structure begin;
and there is also radiative heating as stars, quasars, and AGN's turn on.
Estimates for the average present-day tSZ distortion in a standard
$\Lambda$-CDM cosmology are $\ySZ[t_0]\approx1-2\times10^{-6}$
\cite{ZhangPenTrac}.

One can estimate the spectral distortion by late-time scattering of temperature
anisotropies, by including only the dipole anisotropies from velocities, $\uv$
and neglecting $\udT$.  One should not confuse this effect with the kinetic
Sunyaev-Zel'dovich (kSZ) effect which also is caused by relative velocities of
the baryons and photons: the kSZ effect produces temperature anisotropies not
spectral distortions. The late-time velocity contribution to $\uv$ is given by
\begin{equation}
\label{uVelocity}
\overline{u}_\rmv={2\over3}\int_{t_{\rm rei}}^{t_0} dt\,
{d\overline{\tau}\over dt}\,{v_{\gamma\rm b,rms}^2\over c^2}
\end{equation}
where $t_{\rm rei}$ is the time of the beginning of reionization, $t_0$ is
today, and $v_{\gamma\rm b,rms}$ is the rms baryon velocity wrt to the CMBR
frame.  The calculations is straightforward in a standard $\Lambda$-CDM
cosmology, and is described in \S\ref{app:ComputingVelocities}, obtaining
$\uv\approx3\times10^{-8}$.  There are uncertainties in this number but it is
clear that in the standard cosmology $\udT\ll\uv\ll y_{\rm SZ}$.  However this
was not always the case: the $\uv$ effect is dominated by scattering at
$z\sim\zrei\sim10$, while nearly all of the $\ySZ$ was produced at $z\simlt5$
\cite{ZhangPenTrac}.  So for $z\simgt5$ one expects that $\uv\gg\ySZ$. 
In any case all of these numbers are much less than the current observational
limit $\bar{u},\bar{y}<15\times10^{-6}$ \cite{firas96}.

\subsection{Non-Standard Cosmologies}

If one ventures beyond standard $\Lambda$-CDM cosmologies with Gaussian
inhomogeneities one can imagine that there are regions of the universe where
the $\bfv_{\gamma\rm b}$ is much larger than is observed from our vantage
point.  Spectral distortions, $\uv$, can provide a sensitive probe of regions
of the universe with large velocities such as might occur if there are
non-Gaussian voids.  This is just what is done in ref~\cite{anthro07}.

\section{Discussion}
\label{sec:Discussion}

Scattering of temperature anisotropies inevitably lead to spectral distortions
of the CMBR.  These spectral distortions have been unknown or ignored to date.
They are second order in the amplitude of primordial inhomogeneities and very
small in standard cosmologies; although in standard $\Lambda$-CDM cosmologies
this mechanism was the dominant source of spectral distortions before $z\sim5$.

The main result of this paper is the formalism used to compute this effect both
in the unpolarized case (main text) and for polarized light (appendix).  By
decomposing the spectrum into logarithmic central moments one obtains a
hierarchy of equations where the lower order terms do not depend on the higher
order ones.  One can thus truncate the hierarchy without loss of accuracy.
These are non-perturbative, fully relativistic results for arbitrarily large
spectral distortions and arbitrarily large inhomogeneities.  The most limiting
assumption is that the electrons have small velocity dispersion.  Deviations
from this assumption lead to thermal Sunyaev-Zel'dovich effects which are only
included in an {\it ad hoc} way in this paper.

The spectral distortion in the 0th moment is parameterized by the grayness,
$g$.  It is shown that the scattering of anisotropies, as well as the tSZ
effect, leaves the primordial value $g=0$ unchanged.  The 1st moment is
parametrized by the mean logarithmic temperature $\bar{T}$, which provides a
global (in frequency space) definition of the temperature.

In standard $\Lambda$-CDM cosmologies it is expected that the tSZ effect masks
the distortion caused by the scattering of anisotropies by more than an order
of magnitude, at least in the angular averaged spectral distortion.  It is
possible that spectral distortions described here may never be measured by
looking at anisotropies in the spectral distortion.  Also this effect may allow
one to put limits on variants of standard cosmologies.

The formalism developed here incorporates temperature anisotropies,
polarization, and spectral distortions in a single fairly neat package.  This
might be useful in some pedagogical treatments of the CMBR; especially after
the tSZ effect is incorporated in a less {\it ad hoc} manner.

\subsection{Future Directions}

Here are some directions for future research to extend and apply the results of
this paper
\begin{itemize}
\item check whether in $\Lambda$-CDM cosmologies one can remove the tSZ
contamination by correlating \\ anisotropies in $u$ with anisotropies in
$\bar{T}$.
\item Look for viable cosmological models where the temperature anisotropies
have larger variation than in standard models, leading to larger and perhaps
detectable spectral distortions.  One such case has been done in
ref.~\cite{anthro07}.
\item The formalism used here seems naturally suited to apply to the tSZ
effect, especially relativistic corrections.  It give a simpler path to
expressions for these relativistic corrections, especially in the case of
non-thermal electron distribution functions.
\item Further develop techniques to take the temperature transform of real
spectra.
\item The central moments for scattering of anisotropies as well as the tSZ
effect are small, meaning that the temperature distribution $q[T]$ is narrowly
peaked.  This will not be true of other sources of contamination such as
synchrotron radiation and free-free emission.  One can imaging developing a
new method to filter real spectra, roughly corresponding to a notch filter in
temperature space, that would remove these other contaminants with very high
rejection.
\end{itemize}

\subsection{Other Features}

Here I list some methods and results of this work, which although not the main
focus of the paper, some readers might find more interesting than the main
topic:
\begin{itemize}
\item In \S\ref{app:MomentInverse} is given a general method for inverting
  moments of a broad class of Laplace-like transforms which regularizes
  singular behavior.  The regularization procedure converts the Laplace-like
  transforms into well behaved convolutions which can be inverted using Fourier
  methods.
\item In \S\ref{app:Polarization} a transverse tensor representation
  of the polarization is used , which while equivalent to any other
  representation, leads to extremely simple expressions for the Thomson
  cross-section as well as simple evaluation of angular integrals.
\item In \S\ref{app:Frames} a 3+1, global frame, representation of physical
  quantities is developed and used throughout the paper.  While this is not new
  and may seem less generally covariant, I find it leads to simple and very
  intuitive expressions.
\item In \S\ref{sec:Boltzmann:redshift} a non-perturbative expressions
  for temperature anisotropies (= redshifts) in cosmological space-times w/o
  scattering is given.  These closely resemble the linear theory expressions
  many are already familiar with.
\item In eq.~\ref{ApproximateGrowth} a simple approximation to the
  linear growth of inhomogeneities in a $\Lambda$-CDM cosmology which is
  accurate to better than 1\% is given.
\end{itemize}

\begin{acknowledgments}
I am especially grateful to Robert Caldwell for conversations at the Galileo
Galilei Institute (GGI) for Theoretical Physics which was the seed of this
work. I thank the GGI and Dartmouth College for hospitality during completion
of this work.  This work was partially supported by the INFN at GGI and by the
DoE and the NASA grant NAG 5-10842 at Fermilab.
\end{acknowledgments}

\bibliography{SpectralDynamicsV1_6}

\begin{thebibliography}{20}
\expandafter\ifx\csname natexlab\endcsname\relax\def\natexlab#1{#1}\fi
\expandafter\ifx\csname bibnamefont\endcsname\relax
  \def\bibnamefont#1{#1}\fi
\expandafter\ifx\csname bibfnamefont\endcsname\relax
  \def\bibfnamefont#1{#1}\fi
\expandafter\ifx\csname citenamefont\endcsname\relax
  \def\citenamefont#1{#1}\fi
\expandafter\ifx\csname url\endcsname\relax
  \def\url#1{\texttt{#1}}\fi
\expandafter\ifx\csname urlprefix\endcsname\relax\def\urlprefix{URL }\fi
\providecommand{\bibinfo}[2]{#2}
\providecommand{\eprint}[2][]{\url{#2}}

\bibitem[{\citenamefont{Mather et~al.}(1994)\citenamefont{Mather, Cheng
  et~al.}}]{firas94a}
\bibinfo{author}{\bibfnamefont{J.}~\bibnamefont{Mather}},
  \bibinfo{author}{\bibfnamefont{E.}~\bibnamefont{Cheng}},
  \bibnamefont{et~al.}, \bibinfo{journal}{Ap.~J.}
  \textbf{\bibinfo{volume}{420}}, \bibinfo{pages}{439} (\bibinfo{year}{1994}).

\bibitem[{\citenamefont{Wright et~al.}(1994)\citenamefont{Wright, Mather
  et~al.}}]{firas94b}
\bibinfo{author}{\bibfnamefont{E.}~\bibnamefont{Wright}},
  \bibinfo{author}{\bibfnamefont{J.}~\bibnamefont{Mather}},
  \bibnamefont{et~al.}, \bibinfo{journal}{Ap.~J.}
  \textbf{\bibinfo{volume}{420}}, \bibinfo{pages}{456} (\bibinfo{year}{1994}).

\bibitem[{\citenamefont{Fixsen et~al.}(1996)\citenamefont{Fixsen, Cheng
  et~al.}}]{firas96}
\bibinfo{author}{\bibfnamefont{D.}~\bibnamefont{Fixsen}},
  \bibinfo{author}{\bibfnamefont{E.}~\bibnamefont{Cheng}},
  \bibnamefont{et~al.}, \bibinfo{journal}{Ap.~J.}
  \textbf{\bibinfo{volume}{473}}, \bibinfo{pages}{576} (\bibinfo{year}{1996}).

\bibitem[{\citenamefont{Fixsen and Mather}(2002)}]{firas02}
\bibinfo{author}{\bibfnamefont{D.}~\bibnamefont{Fixsen}} \bibnamefont{and}
  \bibinfo{author}{\bibfnamefont{J.}~\bibnamefont{Mather}},
  \bibinfo{journal}{Ap.~J.} \textbf{\bibinfo{volume}{581}},
  \bibinfo{pages}{817} (\bibinfo{year}{2002}).

\bibitem[{\citenamefont{Birkinshaw}(1999)}]{SZreview}
\bibinfo{author}{\bibfnamefont{M.}~\bibnamefont{Birkinshaw}},
  \bibinfo{journal}{Phys.~Rep.} \textbf{\bibinfo{volume}{310}},
  \bibinfo{pages}{97} (\bibinfo{year}{1999}).

\bibitem[{\citenamefont{Chluba and Sunyaev}(2004)}]{cs04}
\bibinfo{author}{\bibfnamefont{J.}~\bibnamefont{Chluba}} \bibnamefont{and}
  \bibinfo{author}{\bibfnamefont{R.}~\bibnamefont{Sunyaev}},
  \bibinfo{journal}{A\&A} \textbf{\bibinfo{volume}{424}}, \bibinfo{pages}{389}
  (\bibinfo{year}{2004}).

\bibitem[{\citenamefont{Sunyaev and Zel'dovich}(1972)}]{sz72}
\bibinfo{author}{\bibfnamefont{R.}~\bibnamefont{Sunyaev}} \bibnamefont{and}
  \bibinfo{author}{\bibfnamefont{Y.}~\bibnamefont{Zel'dovich}},
  \bibinfo{journal}{Comm.~Ap.} \textbf{\bibinfo{volume}{4}},
  \bibinfo{pages}{301} (\bibinfo{year}{1972}).

\bibitem[{\citenamefont{Chan and Jones}(1975)}]{cj75}
\bibinfo{author}{\bibfnamefont{K.}~\bibnamefont{Chan}} \bibnamefont{and}
  \bibinfo{author}{\bibfnamefont{B.}~\bibnamefont{Jones}},
  \bibinfo{journal}{Ap.~J.} \textbf{\bibinfo{volume}{198}},
  \bibinfo{pages}{245} (\bibinfo{year}{1975}).

\bibitem[{\citenamefont{Salas}(1992)}]{salas92}
\bibinfo{author}{\bibfnamefont{L.}~\bibnamefont{Salas}},
  \bibinfo{journal}{Ap.~J.} \textbf{\bibinfo{volume}{385}},
  \bibinfo{pages}{288} (\bibinfo{year}{1992}).

\bibitem[{\citenamefont{Stebbins}(1997{\natexlab{a}})}]{stebbins97a}
\bibinfo{author}{\bibfnamefont{A.}~\bibnamefont{Stebbins}}, in
  \emph{\bibinfo{booktitle}{The Cosmic Microwave Background}}, edited by
  \bibinfo{editor}{\bibfnamefont{C.}~\bibnamefont{Lineweaver}},
  \bibinfo{editor}{\bibfnamefont{J.}~\bibnamefont{Bartlett}},
  \bibinfo{editor}{\bibfnamefont{A.}~\bibnamefont{Blanchard}},
  \bibinfo{editor}{\bibfnamefont{M.}~\bibnamefont{Signore}}, \bibnamefont{and}
  \bibinfo{editor}{\bibfnamefont{J.}~\bibnamefont{Silk}}
  (\bibinfo{publisher}{Kluwer Academic}, \bibinfo{year}{1997}{\natexlab{a}}),
  vol. \bibinfo{volume}{502} of \emph{\bibinfo{series}{NATO ASI}}, pp.
  \bibinfo{pages}{241--270}.

\bibitem[{\citenamefont{Stebbins}(1997{\natexlab{b}})}]{stebbins97b}
\bibinfo{author}{\bibfnamefont{A.}~\bibnamefont{Stebbins}}
  (\bibinfo{year}{1997}{\natexlab{b}}), \eprint{astroph/9709065}.

\bibitem[{\citenamefont{Challinor and Lasenby}(1998)}]{cl98}
\bibinfo{author}{\bibfnamefont{A.}~\bibnamefont{Challinor}} \bibnamefont{and}
  \bibinfo{author}{\bibfnamefont{A.}~\bibnamefont{Lasenby}},
  \bibinfo{journal}{Ap.~J.} \textbf{\bibinfo{volume}{198}}, \bibinfo{pages}{1}
  (\bibinfo{year}{1998}).

\bibitem[{\citenamefont{Itoh et~al.}(1998)\citenamefont{Itoh, Kohyama, and
  Nozawa}}]{ikn98}
\bibinfo{author}{\bibfnamefont{N.}~\bibnamefont{Itoh}},
  \bibinfo{author}{\bibfnamefont{Y.}~\bibnamefont{Kohyama}}, \bibnamefont{and}
  \bibinfo{author}{\bibfnamefont{S.}~\bibnamefont{Nozawa}},
  \bibinfo{journal}{Ap.~J.} \textbf{\bibinfo{volume}{502}}, \bibinfo{pages}{7}
  (\bibinfo{year}{1998}).

\bibitem[{\citenamefont{Zaldarriaga and Seljak}(2000)}]{CMBFAST}
\bibinfo{author}{\bibfnamefont{M.}~\bibnamefont{Zaldarriaga}} \bibnamefont{and}
  \bibinfo{author}{\bibfnamefont{U.}~\bibnamefont{Seljak}},
  \bibinfo{journal}{Ap.~J.~Suppl.} \textbf{\bibinfo{volume}{129}},
  \bibinfo{pages}{431} (\bibinfo{year}{2000}).

\bibitem[{\citenamefont{Zhang et~al.}(2004)\citenamefont{Zhang, Pen, and
  Trac}}]{ZhangPenTrac}
\bibinfo{author}{\bibfnamefont{P.}~\bibnamefont{Zhang}},
  \bibinfo{author}{\bibfnamefont{U.}~\bibnamefont{Pen}}, \bibnamefont{and}
  \bibinfo{author}{\bibfnamefont{H.}~\bibnamefont{Trac}},
  \bibinfo{journal}{M.N.R.A.S.} \textbf{\bibinfo{volume}{347}},
  \bibinfo{pages}{1224} (\bibinfo{year}{2004}).

\bibitem[{\citenamefont{Caldwell and Stebbins}(2007)}]{anthro07}
\bibinfo{author}{\bibfnamefont{R.}~\bibnamefont{Caldwell}} \bibnamefont{and}
  \bibinfo{author}{\bibfnamefont{A.}~\bibnamefont{Stebbins}}
  (\bibinfo{year}{2007}), \bibinfo{note}{in preparation}.

\bibitem[{\citenamefont{Peacock}(1999)}]{p99}
\bibinfo{author}{\bibfnamefont{J.}~\bibnamefont{Peacock}}
  (\bibinfo{year}{1999}).

\bibitem[{\citenamefont{Bardeen et~al.}(1986)\citenamefont{Bardeen, Bond,
  Kaiser, and Szalay}}]{BBKS}
\bibinfo{author}{\bibfnamefont{J.}~\bibnamefont{Bardeen}},
  \bibinfo{author}{\bibfnamefont{R.}~\bibnamefont{Bond}},
  \bibinfo{author}{\bibfnamefont{N.}~\bibnamefont{Kaiser}}, \bibnamefont{and}
  \bibinfo{author}{\bibfnamefont{A.}~\bibnamefont{Szalay}},
  \bibinfo{journal}{Ap.~J.} \textbf{\bibinfo{volume}{304}}, \bibinfo{pages}{15}
  (\bibinfo{year}{1986}).

\bibitem[{\citenamefont{Tegmark et~al.}(2006)\citenamefont{Tegmark, Eisenstein
  et~al.}}]{SDSSlrg}
\bibinfo{author}{\bibfnamefont{M.}~\bibnamefont{Tegmark}},
  \bibinfo{author}{\bibfnamefont{D.}~\bibnamefont{Eisenstein}},
  \bibnamefont{et~al.}, \bibinfo{journal}{Phys.~Rev.~D}
  \textbf{\bibinfo{volume}{74}}, \bibinfo{pages}{123507}
  (\bibinfo{year}{2006}).

\bibitem[{\citenamefont{Spergel et~al.}(1996)\citenamefont{Spergel, Bean
  et~al.}}]{WMAP}
\bibinfo{author}{\bibfnamefont{D.}~\bibnamefont{Spergel}},
  \bibinfo{author}{\bibfnamefont{R.}~\bibnamefont{Bean}}, \bibnamefont{et~al.}
  (\bibinfo{year}{1996}), \eprint{astro-ph/0603449}.

\end{thebibliography}

\appendix

\section{Logarithmic Moments From Brightness}
\label{app:MomentInverse}

In this paper the quantities $g$, $\bar{T}$, and $u_{(n)}$ are used.  These are
defined in terms of the $\eta_{(n)}$ in 
eq.s~(\ref{grayness},\ref{meantemp},\ref{centralmoments}), which in turn are
defined in terms of the temperature transform $q[\ln[T],\cdots]$.  We will
not specify how to go from a normal brightness spectrum to $q$, but we now
specify how to get the $\eta_{(n)}$ from the spectrum.  

Let us begin with some mathematical preliminaries.  Suppose we have a
convolution of the form
\begin{equation}
f[x]=\int_{-\infty}^\infty dy\,k[x-y]\,\tilde{f}[y].
\label{transform}
\end{equation}
where the convolution kernel, $k$, is a known function.  The relation between
the moments of $f$ and the moments of $\tilde{f}$ is complicated if $k$ does
not go to zero sufficiently rapidly for large positive and negative argument.
In particular if
\begin{eqnarray}
\lim_{z\rightarrow+\infty}k[z]=\sum_{i=1}^{n_+}a_i e^{\alpha_i z} \\ \nonumber
\lim_{z\rightarrow-\infty}k[z]=\sum_{j=1}^{n_-}b_j e^{\beta_j  z}
\end{eqnarray}
where the $\alpha_i$ are non-negative and the $\beta_i$ are non-positive this
presents a complication.  

To proceed one can define the differential operator
\begin{equation}
\hat{D}_z=\prod_{i=1}^{n_+}\left({d\over dz}-\alpha_i\right)
          \prod_{j=1}^{n_-}\left({d\over dz}- \beta_j\right)\ ,
\end{equation}
so that $\hat{D}_z k[z]$ goes to zero rapidly for both large positive and large
negative argument.  On the other hand if $k$ does fall off rapidly one can set
$\hat{D}_z$ to 1.

Define the moments
\begin{eqnarray}
       \kappa_l &=&\int_{-\infty}^\infty dz\,z^l \hat{D}_z k[z] \\ \nonumber
        \zeta_n &=&\int_{-\infty}^\infty dx\,x^n \hat{D}_x f[x] \\ \nonumber
\tilde{\zeta}_m &=&\int_{-\infty}^\infty dy\,y^m   \tilde{f}[y] 
\label{integrals}
\end{eqnarray}
where the $\kappa_l$ are known in the sense they can be computed numerically if
not analytically.  The $\zeta_n$ and $\tilde{\zeta}_m$ are related by
\begin{equation}
\zeta_n     =\sum_{m=0}^n {n!\,\kappa_{n-m}\,\tilde{\zeta}_m\over m!\,(n-m)!}
\qquad
\tilde{\zeta}_m
            =\sum_{n=0}^m {m!\,\tilde{\kappa}_{m-n}\,\zeta_n\over n!\,(m-n)!}
\label{zetamoments}
\end{equation}
where the $\tilde{\kappa}_k$ can be solved for recursively:
\begin{equation}
\tilde{\kappa}_k={k!\over\kappa_0}\left(\delta_{k,0}
-\sum_{l=0}^{k-1}{\kappa_{k-l}\over(k-l)!}\,{\tilde{\kappa}_l\over l!}\right)
\ .
\end{equation}
The inverse exists if and only if $\kappa_0\ne0$.

Now let us apply this to the temperature transform,
eq.~(\ref{temperaturetransform}) which is of the form of eq.~(\ref{transform}),
when one uses 
$x=\ln[{h\nu\over\kB}]$,  $y=\ln[T]$, $f=n$ and $\tilde{f}=q$.  Thus one finds
$k[z]=(e^{e^z}-1)^{-1}$ and $\tilde{\zeta}_m=\eta_{(m)}$.  The function $k[z]$
does go to zero for large positive argument so $n_+=0$ but
$\lim_{z\rightarrow-\infty}k[z]=e^{-z}-{1\over2}$ so $n_-=2$ with $\beta_1=-1$
and $\beta_2=0$.  Thus the differential operator needed to regulate divergences
is $\hat{D}_z={d^2\over dz^2}+{d\over dz}$ and
\begin{equation}
\hat{D}_z k[z]=
{e^z\, e^{e^z}\,(e^z\,(e^{e^z}+1)-2\,(e^{e^z}-1))\over(e^{e^z}-1)^3}
.\end{equation}
With this form, one can show that for large $l$ the $\kappa_l$ integral is
dominated by large negative argument and one can show
\begin{equation}
\lim_{l\rightarrow\infty}\kappa_l={1\over6}\int_0^1 dx\,\ln[x]^l
={1\over6}(-1)^l\,l!
\label{kappalimit}
\end{equation}
and this limit is approached fairly rapidly.  In table~\ref{table:kappa}
the first few numerical for the $\kappa_n$ and $\tilde{\kappa}_m$ are given. 

\begin{table}[t]
\begin{tabular}{| l | l | }              \hline
\hskip34pt      $\kappa_0=+{1\over2}$ & 
        $\tilde{\kappa}_0=+2        $ \\ \hline
$(-1)^1{6\over1!}\kappa_1=-0.7820   $ &
        $\tilde{\kappa}_1=-0.52132  $ \\ \hline
$(-1)^2{6\over2!}\kappa_2=+2.0217   $ &
        $\tilde{\kappa}_2=-2.42382  $ \\ \hline
$(-1)^3{6\over3!}\kappa_3=+0.5424   $ &
        $\tilde{\kappa}_3=+6.17299  $ \\ \hline
$(-1)^4{6\over4!}\kappa_4=+1.1669   $ & 
        $\tilde{\kappa}_4=-7.76751  $ \\ \hline
$(-1)^5{6\over5!}\kappa_5=+0.9473   $ &
        $\tilde{\kappa}_5=+0.74497  $ \\ \hline
$(-1)^6{6\over6!}\kappa_6=+1.0145   $ &
        $\tilde{\kappa}_6=+23.7175  $ \\ \hline
$(-1)^7{6\over7!}\kappa_7=+0.9964   $ &
        $\tilde{\kappa}_7=-68.4985  $ \\ \hline
$(-1)^8{6\over8!}\kappa_8=+1.0008   $ &
        $\tilde{\kappa}_8=+108.645  $ \\ \hline
$(-1)^9{6\over9!}\kappa_9=+0.9999   $ &
        $\tilde{\kappa}_9=-55.2136  $ \\ \hline
\end{tabular}
\caption{Listed are the coefficients $\kappa_{(k)}$ and $\tilde{\kappa}_{(k)}$
used in  eq.~(\ref{zetamoments}) computed for $k\le9$ for a temperature
transform.  The $\kappa_{(k)}$ are computed using the integral in
eq.~(\ref{integrals}) (analytically for $k=0$ but numerically otherwise) and
expressed divided by the asymptotic form of eq.~(\ref{kappalimit}).  The
$\tilde{\kappa}_{(k)}$ are computed from the $\kappa_{(k)}$ using a backwards
substitution algorithm.}
\label{table:kappa}
\end{table}

To go from brightness to $\eta_{(n)}$ for $n\le N$ first compute the occupation
number $n[\nu]=c^2 I_\nu/(2h\nu^3)$, then compute the moments
\begin{equation}
\zeta_n=
\int_0^\infty {d\nu\over\nu}\,\ln[{h\nu\over\kB}]^n 
{\partial\over\partial\nu}\left(\nu^2{\partial n\over\partial\nu}\right)
\end{equation}
for $n\le N$.  Then use eq.~(\ref{zetamoments}) and $\tilde{\kappa}_l$ from
table~\ref{table:kappa} to compute $\tilde{\zeta}_m=\eta_{(m)}$.

\section{Global Frames}
\label{app:Frames}

In this paper the occupation number as $n[\nu,\hatbfc,\bfx,t]$ is used, and the
notation indicates a specific spatial coordinates $\bfx$, and temporal
coordinates $t$, has been chosen, as well as a specific rest-frame in which
$\nu$ and the direction of travel $\hatbfc$ is measured throughout space-time.
This represents a particular choice of a global 4-velocity field for the
``observer''.  Denote the 4-velocity field by $u^\alpha$, which is normalized
$u^\alpha g_{\alpha\beta} u^\beta=-1$ where $g_{\alpha\beta}$ is the metric.
The 4-velocity space-time gradients are canonically decomposed into an
expansion rate, a shear tensor, a vorticity vector, and a proper 4-acceleration
which are respectively
\begin{eqnarray}
\theta&=&{u^\alpha}_{;\alpha} \\ \nonumber
\sigma_{\alpha\beta}&=&{1\over2}{P_\alpha}^{\gamma}\,
\left(u_{\gamma;\delta}+u_{\delta;\gamma}-{2\over3}\theta\,P_{\delta\gamma}
\right)\,{P^\delta}_\beta \\ \nonumber
\omega^\alpha&=&-\epsilon^{\alpha\beta\gamma\delta}u_{\beta;\gamma}u_\delta\\
\nonumber
a_\alpha&=&u^\beta u_{\alpha;\beta}\ .
\end{eqnarray}
Here $P_{\alpha\beta}=g_{\alpha\beta}+u_{\alpha\beta}$ is the spatial
projection tensor and $\epsilon_{\alpha\beta\gamma\delta}$ is the 4-d
Levi-Civita tensor.  Thus 
\begin{equation}
u_{\alpha;\beta}={1\over3}\theta\,P_{\alpha\beta}+\sigma_{\alpha\beta}
+\omega^\gamma u^\delta\epsilon_{\alpha\beta\gamma\delta}-a_\alpha u_\beta\ .
\end{equation}
Note that $P_{\alpha\beta}$ is involutive, \ie
${P^\alpha}_\beta\,{P^\beta}_\gamma={P^\alpha}_\gamma$, both $P_{\alpha\beta}$
and $\sigma_{\alpha\beta}$ are symmetric, and $\sigma_{\alpha\beta}$,
$\omega^\gamma$, and $a_\alpha$ are purely spatial, \ie
$\sigma_{\alpha\beta}u^\beta=\omega_\alpha u^\alpha=a_\alpha u^\alpha=0$.

The tensors $P_{\alpha\beta}$ and $\epsilon_{\alpha\beta\gamma\delta}$ allow
one to perform 3-d vector analysis in the frame $u^\alpha$, \eg
\begin{equation}
\bfa{\bf\cdot}\bfb=a^\alpha P_{\alpha\beta}b^\beta \qquad
(\bfa{\bf\times}\bfb){\bf\cdot}\bfc
=a^\alpha b^\beta c^\gamma\epsilon_{\alpha\beta\gamma\delta}u^\delta\ .
.\end{equation}

In the geometric optics limit photons will follow null geodesics and the
frequency observed in a particular frame is $\nu[\lambda]\propto
u_\alpha{dx^\alpha\over d\lambda}$ where $\lambda$ is an affine parameter.
Alternately one can define the redshift for that geodesic, $z[\lambda]$, such
that for all photons following that geodesic $\nu\propto1+z[\lambda]$.  A frame
also induces a time parameterization on each geodesic:
$t[\lambda]=-c\int^\lambda d\lambda u_\alpha {dx^\alpha\over d\lambda}$, and
one will need a convective derivative, \ie the derivative of any quantity along
the geodesic, denoted by ${D\over Dt}$.  In any frame
\begin{equation}
{D\ln[\nu]\over Dt}={D\ln[1+z]\over Dt}
=-{1\over3}\theta-\hatbfc\cdot\sigma\cdot\hatbfc+\hatbfc\cdot\bfa
\end{equation}
where
\begin{equation}
\hatbfc\cdot{\bf \sigma}\cdot\hatbfc=
{{dx^\alpha\over d\lambda}\sigma_{\alpha\beta}{dx^\beta\over d\lambda}
 \over {dx^\gamma\over d\lambda}P_{\gamma\delta}{dx^\delta\over d\lambda}}
\qquad
\hatbfc\cdot\bfa=
{{dx^\alpha\over d\lambda}\cdot a_\alpha\over
\sqrt{{dx^\gamma\over d\lambda}P_{\gamma\delta}{dx^\delta\over d\lambda}}}\ .
\end{equation}
Note that if for a frame $\omega^\gamma=0$ then the 3-d velocity gradient
tensor is $\nabla\cdot\bfv={1\over3}\theta\,\bfI+{\bf\sigma}$.

\section{Polarization}
\label{app:Polarization}

The description of the photon distribution in terms of $n$ is incomplete
because a beam a photons can be polarized.  Quantitatively this is not liable
to have a large effect on the evolution of $n$ but it does have some effect.
Furthermore to predict the spectral distortions of different polarization modes
one will of course need to include it in the equations.  The equations in this
appendix are not actually used in the paper but we include them for reference.

\subsection{The Polarization Tensor}

A beam of photons traveling in direction $\hatbfc$ and space-time point${}^a$
can be described by four Stokes parameters as a function of frequency $I_\nu$,
$Q_\nu$, $U_\nu$, and $V_\nu$. Here $I_\nu$ is the intensity used above,
$Q_\nu$ and $U_\nu$ parameterize the linear polarization and $V_\nu$ the
circular polarization.  In any particular frame, these four quantities can be
used to define a 3-d, rank 2 real tensor $\Pt$ called the {\it polarization
tensor}.  Here the notation $\Tt$ is used to indicate a 3-d tensor, usually of
rank 2, which is transverse, \ie
$\hatbfc\cdot\Tt[\hatbfc]=\Tt[\hatbfc]\cdot\hatbfc={\bf0}$. The simplest such
tensor is the {\it transverse tensor} defined by
\begin{equation}
\label{TransverseTensor}
\It[\hatbfc]\equiv\bfI-\hatbfc\otimes\hatbfc
\end{equation}
which is the unique transverse involutive ($\It\cdot\It=\It$) rank 2 tensor,
and can be used to project into the space perpendicular to $\hatbfc$, \eg
$\overleftrightarrow{\bfT}=\It\cdot\bfT\cdot\It$.

\begin{figure}[b]
\footnotetext[1]{One doesn't mean a literal point, but some finite if small
region.  One needs a finite space-time region dictated by the uncertainty
principles $\Delta t\,\Delta\nu\simlt1$, and
$\nu\,\delta\hatbfc\,|\Delta\bfx|\simlt c$ to obtain good angular and frequency
resolution.  The size of astronomical telescopes are often dictated by these
requirements.}
\end{figure}

To construct $\Pt$ one needs to define, for each $\hatbfc$, two direction
vectors $\pvector{i}[\hatbfc]$ for $i=1,2$ which are transverse
($\hatbfc\cdot\pvector{i}=0$), orthonormal
$\pvector{i}\cdot\pvector{j}=\delta_{ij}$, and have handedness given by
$\pvector{1}\times\pvector{2}=\hatbfc$.  The linear polarization Stokes
parameters $Q_\nu$ and $U_\nu$ are defined up to a rotation about the $\hatbfc$
axis, and they are chosen such that electric field oscillations in the
$\pm\pvector{1}$ direction corresponds to a $Q_\nu>0$, while electric field
oscillations in the $\pm(\pvector{1}+\pvector{2})/\sqrt{2}$ direction
corresponds corresponds to $U_\nu>0$.  In this case $\Pt$ is
\begin{eqnarray}
\Pt[\nu,\hatbfc,\bfx,t]&=&
   {1\over2}I_\nu\,\left( \pvector{1}\otimes\pvector{1}
                         +\pvector{2}\otimes\pvector{2}\right) \\\nonumber
&+&{1\over2}Q_\nu\,\left( \pvector{1}\otimes\pvector{1}
                         -\pvector{2}\otimes\pvector{2}\right) \\\nonumber
&+&{1\over2}U_\nu\,\left( \pvector{1}\otimes\pvector{2}
                         +\pvector{2}\otimes\pvector{1}\right) \\\nonumber
&+&{1\over2}V_\nu\,\left( \pvector{1}\otimes\pvector{2}
                         -\pvector{2}\otimes\pvector{1}\right)\ ,
\end{eqnarray}
where $\otimes$ indicates an outer product.  For most purposes $\Pt$
provides a sufficient description of the radiation field in astronomical
applications.  The three rotational invariants are
\begin{eqnarray}
I_\nu&=&\Tr[\Pt]              \\ \nonumber
V_\nu&=&\Tr[\hatbfc\times\Pt] \\ \nonumber
\sqrt{Q_\nu^2+U_\nu^2}&=&
\sqrt{2\Tr[\Pt\cdot\Pt]-\Tr[\Pt]^2+\Tr[\hatbfc\times\Pt]^2}
\end{eqnarray}
which are, respectively, the intensity, the circular polarization, and the
amplitude of linear polarization.  Unpolarized light has $Q_\nu=U_\nu=V_\nu=0$
so
\begin{equation}
\label{unpolarized}
\Pt[\hatbfc\,\cdots]={1\over2}I_\nu[\hatbfc,\cdots]\,\It[\hatbfc]\ .
\end{equation}
since $\It=\pvector{1}\otimes\pvector{1}+\pvector{2}\otimes\pvector{2}$.

\subsection{The Occupation Number Tensor}

From $\Pt$ one can construct a quantum mechanical occupation number
tensor
\begin{equation}
\label{OccupationTensor}
\Nt[\nu,\hatbfc,\bfx,t]={c^2\over2h\nu^3}\,
\Pt[\nu,\hatbfc,\bfx,t]
\end{equation}
which is dimensionless.  The occupation number used in the main text is
$n[\nu,\hatbfc,\bfx,t]=\Tr[\Nt[\nu,\hatbfc,\bfx,t]]$.  True blackbody radiation
is unpolarized and has
\begin{equation}
\label{BlackbodyPolarization}
\Nt^{\rm BB}[\nu,\cdots]={1\over2}{\It\over e^{h\nu\over\kB T}-1}\ .
\end{equation}
where $T$ is the temperature

\subsection{Tensor Temperature Transform, Moments, Fokker-Planck Expansion}

In analogy with eq.s~(\ref{temperaturetransform},\ref{logmoments}) one can
define the tensor temperature transform tensor by
\begin{equation}
\label{tensortemperaturetransform}
\Nt[\nu,\cdots]
=\int_{-\infty}^\infty d\ln[T]\,
{\Qt[\ln[T],\cdots]\,\over e^{h\nu\over\kB T}-1}
\end{equation}
and the tensor logarithmic moments by
\begin{equation}
\label{tensorlogmoments}
\Ht{n}[\cdots]\equiv
\int_{-\infty}^\infty d\ln[T]\,\ln[T]^n\,\Qt[\ln[T],\cdots]\ .
\end{equation}
so that $\eta_{(n)}=\Tr[\Ht{n}]$, $g=\Tr[\Ht{0}]$ and
$\bar{T}=e^{\Tr[\Ht{1}]/\Tr[\Ht{0}]}$ (see
eq.s~(\ref{grayness},\ref{meantemp})). Finally in analogy with
eq.~(\ref{tensorcentralmoments}) define
\begin{equation}
\Ut{n}\equiv\sum_{k=0}^n {(-1)^kn!\over k!\,(n-k)!}\,
{\Ht{k}\over1-g}\,\ln[\bar{T}]^{n-k}
\label{tensorcentralmoments}
\end{equation}
so $u_{(n)}=\Tr[\Ut{n}]$; and by definition $\Tr[\Ut{0}]=1$ and
and $\Tr[\Ut{1}]=0$.  The Fokker-Planck series corresponding to
eq.s~(\ref{FokkerPlanck},\ref{FokkerPlanckAlt},\ref{FokkerPlanckCoeffs}) are 
\begin{eqnarray}
\Nt[\nu,\cdots]
&=&       \sum_{m=0}^\infty{\Dt_{(m)}[\cdots]\over m!}
                      \,\dn{m}\left[{h\nu\over\kB T_0   }\right] \\ \nonumber
&=&(1-g)\,\sum_{n=0}^\infty{\Ut{n}[\cdots]\over n!}
                      \,\dn{n}\left[{h\nu\over\kB\bar{T}}\right]\ .
\label{TensorFokkerPlanck}
\end{eqnarray}
where
\begin{eqnarray}
{\Dt_{(m)}\over1-g}&=&\sum_{n=0}^m{m!\over n!\,(m-n)!}\,
\ln\left[{\bar{T}\over T_0}\right]^{m-n}\,\Ut{n} \\ \nonumber
\Ut{n}&=&\sum_{m=0}^n{n!\over m!\,(n-m)!}\,
\ln\left[{T_0\over\bar{T}}\right]^{n-m}\,{\Dt_{(m)}\over1-g}
\label{TensorFokkerPlanckCoeffs}
\end{eqnarray}
and $T_0$ is an arbitrary reference temperature.

\subsection{The Tensor Boltzmann Equation}

In vacuum (\ie without scattering), in the geometric optics limit,
$\Pt$ is conserved along geodesics once one takes into account
redshifting (changes in $\nu$) and any rotation of linear polarization (mixing
of $Q$ and $U$ which preserves $Q^2+U^2$).  All this can be absorbed into a
convective derivative.  Define a ``basic'' convective derivative for transverse
tensors, similar to that in eq.~(\ref{convective}):
\begin{eqnarray}
{\calD\over\calD t}\Tt&=&
{\partial\over\partial t}\Tt
+{d\bfx\over dt}\cdot\nabla_\bfx\Tt
+{d\hatbfc\over dt}\cdot\nabla_\hatbfc\Tt \\ \nonumber
&&\hskip28pt+{d\varphi\over dt}\left(\Tt\times\hatbfc-\hatbfc\times\Tt\right)
\end{eqnarray}
where $d\varphi/dt$ is the rate of rotation (if any) of the linear
polarization.  This last term does not effect the intensity or the circular
polarization.  The full convective derivative, $D/Dt$, includes redshifting,
but depends on the context.  In frequency space
\begin{equation}
{D\over Dt}={\calD\over\calD t}
+{d\ln[1+z]\over dt}{\partial\over\partial\ln[\nu]}
\end{equation}
in temperature space
\begin{equation}
{D\over Dt}={\calD\over\calD t}
+{d\ln[1+z]\over dt}{\partial\over\partial\ln[T]}
\end{equation}
for the logarithmic moment tensor
\begin{equation}
{D\over Dt}\Ht{n}={\calD\over\calD t}\Ht{n}-n\,{d\ln[1+z]\over dt}\,\Ht{n-1}
\end{equation}
for $\ln[\bar{T}]$
\begin{equation}
{D\over Dt}\ln[\bar{T}]={\calD\over\calD t}\ln[\bar{T}]-{d\ln[1+z]\over dt}
={\calD\over\calD t}\ln\left[{\bar{T}\over1+z}\right]
\end{equation}
while $D/Dt=\calD/\calD t$ for $g$ and $\Ut{n}$.

Non-convective effects include scattering, refraction, dispersion, etc.; but in
cosmological application the most important effect is (Thomson) scattering off
of free electrons.  Before proceeding with Thomson scattering define the
angular average operator
\begin{equation}
\label{AngleAverage}
\AngleAverage{F[a,b']}\equiv
{1\over4\pi}\int d^2\hatbfc'\,F[a[\hatbfc],b[\hatbfc']]\ .
\end{equation}
Note that the convolution operator of eq.~(\ref{SigmaOperator}) is
\begin{eqnarray}
\SigmaOp  F[a,b']&=&{3\over2}
\AngleAverage{F[a,b']\,{1+(\hatbfc\cdot\hatbfc')^2\over2}} \\ \nonumber
&=&{1\over2}\AngleAverage{F[a,b']\,\Tr[\It\cdot\It'\cdot\It]}
\end{eqnarray}
which is where $\SigmaOp$ comes from.

When one includes Thomson scattering off of cold electrons the collisional
Boltzmann equation becomes
\begin{equation}
\label{TensorBoltzmann}
{D\over Dt}\Nt=
{d\tau\over dt}\,\left(-\Nt+\It\cdot\AngleAverage{\Nt'}\cdot\It\right)
\end{equation}
where $\tau$ is the Thomson optical depth as defined in
\S\ref{sec:Spectrum:SZ}.  The first term on the rhs, $-\Nt$, gives the light
scattered out of the beam, while the 2nd term is the light scattered into the
beam.  Clearly the scattered light is transverse, as it must be.  It is also
clear from this equation that the scattering of the antisymmetric part of the
polarization tensor does not couple to the symmetric part.  This means that
circular polarization evolves independently of linear polarization and
intensity.

Note that even if the incoming light is unpolarized the outgoing light will, in
general be linearly polarized, \ie scattered unpolarized light will be
polarized.  The approximation used in the main text is to use unpolarized
light in scattering term $\Nt[\nu,\hatbfc,\cdots]={1\over2}
n[\nu,\hatbfc,\cdots]\,\It[\hatbfc]$ and then take the trace to get scattered
occupation number, so that
\begin{equation}
\label{unpolarizedBoltzmann}
{Dn\over Dt}={d\tau\over dt}
{3\over16\pi}\int d^2\hatbfc'\,\left(1+(\hatbfc\cdot\hatbfc')^2\right)\,
\left(n'-n\right)
\end{equation}
which is the origin of eq.~(\ref{UnpolarizedCrossSection}).  This is correct if
there is only one scattering of initially unpolarized light, but not for
multiple scattering.  In many applications the single scattering approximation
is good.  In any case one knows that the CMBR is at most about 10\% linearly
polarized so the approximate equation is not liable to lead to large errors.

From eq.~(\ref{TensorBoltzmann}) one can derive the Boltzmann equation for $g$,
$\ln[\bar{T}]$ and $\Ut{n}$:
\begin{eqnarray}\label{TensorBoltzmannCentral} \nonumber
{\calD\ln[1-g]\over\calD t}&=&{d\tau\over dt}\,
\left(1-{3\over2}\Tr\left[\It\cdot\bfA_{(0,0)}\cdot\It\right]\right)
\\ \nonumber
{\calD\over\calD t}\ln\left[{\bar{T}\over1+z}\right]&=&
{3\over2}{d\tau\over dt}\,
\Tr\left[\It\cdot\left(\bfA_{(1,0)}+\bfA_{(0,1)}\right)\cdot\It\right]
\\ \nonumber
{\calD\over\calD t}\Ut{n}&=&
{3\over2}{d\tau\over dt}\,\Biggl(
\Tr\left[\It\cdot\bfA_{(0,0)}\cdot\It\right]\,\Ut{n}\\ \nonumber
&&\hskip-45pt-n\,\Tr\left[
\It\cdot\left(\bfA_{(1,0)}+\bfA_{(0,1)}\right)\cdot\It\right]\,\Ut{n-1} \\
&&\hskip-45pt+\sum_{m=0}^n{n!\over m!(n-m)!}
\It\hskip-2pt\cdot\hskip-2pt\bfA_{(n-m,m)}\hskip-2pt\cdot\hskip-2pt\It
\Biggr)
\end{eqnarray}
where 
\begin{equation}
\label{Atensor}
\bfA_{(k,l)}\equiv
\AngleAverage{{1-g'\over1-g}\,\ln\left[{\bar{T}'\over\bar{T}}\right]^k\Ut{l}'}
\ .
\end{equation}
If one sets $\Ut{n}=u_{(n)}\It/2$ on the rhs and take the trace one will
recover eq.~(\ref{CentralBoltzmann}).

\subsubsection{Neither Gray nor Circular}

One sees from eq.s~(\ref{TensorBoltzmannCentral}) that if $\Ut{n}$ is initially
symmetric it will remain so.  So under the assumption of Thomson scattering by
cold electrons one sees that no circular polarization will develop.
Furthermore one sees that if one sets $g=1$ and $\Ut{0}=\It/2$, which is the
expected thermal initial conditions, then
\begin{equation}
{\calD g\over\calD t}=0 \qquad\qquad {\calD\over\calD t}\Ut{0}={\bf 0}\ .
\end{equation}
So as with the unpolarized equations there are no grayness terms; and the
spectrum of the polarization terms are given by derivatives of a blackbody, not
a blackbody itself.  Thus one can simply set the circular polarization and
grayness to zero.  Of course this may not remain true if other radiative
processes are included.

\subsubsection{Perturbative Analysis}

Note that the perturbative analysis of \S~\ref{sec:SpectralDynamics:Perturb} is
unchanged from the unpolarized version of the equations, namely 
\begin{eqnarray} \nonumber
                           g\sim\Ut{0}&\sim&\calO[\infty]       \\ \nonumber
\ln\left[{\bar{T}'\over\bar{T}}\right]&\sim&\calO[1]            \\
                                \Ut{n}&\sim&\calO[n] \qquad n\ge1\ .
\end{eqnarray}

\subsubsection{Numerical Scaling}

One could imagine numerically simulating eq.s~(\ref{TensorBoltzmannCentral})
over some space-time volume.  Without scattering the full radiative transfer
computation scales like the number of space-time points ($\#_{\rm x}^3\#_{\rm
t}$) times the number of frequencies ($\#_\nu$), times the number of angular
resolution elements ($\#_\theta^2$).  In general scattering can increase this
by additional factors of $\#_\nu \#_\theta^2$. However by using the moment
decomposition one can reduce $\#_\nu$ to a relatively small number of moments
$\#_m$, \eg $\#_m=1$ if one only wants to simulate $\bar{T}$.  Within the
context of Thomson scattering by cold electrons one does this without
approximation.  Furthermore the scattering term only adds an additional factor
of $\#_m^2$.  Thus the total computational scaling is
$\#_x^3\#_t\#_\theta^2\#_m^3$.  Of course full relativistic radiative transfer
requires the timestep to be less than the light crossing time of the spatial
resolution element.  For non-relativistic flows various quasi-static
approximations can be used to reduce the number of timesteps.

To achieve the $\#_\theta^2\#_m^3$ scaling one can use the relation
\begin{equation}
\label{AfromB}
\bfA_{(k,l)}[\hatbfc,\cdots]={1\over1-g}
\sum_{i=1}^k{(-1)^i k!\over i!(k-i)!}\ln[\bar{T}[\hatbfc,\cdots]]^i
\bfB_{(k-i,l)}[\cdots]
\end{equation}
where
\begin{equation}
\label{Btensor}
\bfB_{(k,l)}\equiv\AngleAverage{(1-g')\,\ln\left[\bar{T}'\right]^k\Ut{l}'}\ .
\end{equation}
The important point is that $\bfB_{(k,l)}$ is independent of $\hatbfc$ and
depends only $\bfx$ and $t$.  One never has to store or compute functions of
both $\hatbfc$ and $\hatbfc'$.  For each space-time point, to compute all the
$\bfB_{(l,k)}$'s scales like $\#_\theta^2\#_m^2$.  One can then compute the
$\bfA_{(l,k)}$ locally at each $\hatbfc$ element, using eq.~(\ref{AfromB}),
which scales like $\#_m^3$.  Then computing the scattering terms in
eq.~(\ref{TensorBoltzmannCentral}) is also local in $\hatbfc$ space, scaling as
$\#_m^2$.  Finally the convective derivative is quasi-local in $\bfx$, $t$,
$\hatbfc$-space, involving only adjacent points.  For large $\#_m$ the largest
scaling is to compute eq.~(\ref{AfromB}).

\subsection{Gravitational Field of the Spectral Distortions}

An additional complication in truncating the moment distribution is that the
stress-energy of the photons does not involve only the lowest order moments,
$\Ut{n}$.  So even though in the radiative transfer sector the moment
truncation is exact and non-perturbative, a truncation of the spectral
distortion can lead to only approximate information about the gravitational
field of the photons.  In many cases however the gravitational field of the
photons, let alone the spectral distortions is completely negligible. 

\section{Formulae for Late-Time Spectral Distortion}
\label{app:ComputingVelocities}

In practice one will find it convenient to express the eq.~(\ref{uVelocity}) as
an integral over cosmological redshift, $z$, rather than $t$, and one can use
the relationship
\begin{equation}
{dln[1+z]\over dt}=-H[z]
\end{equation}
where $H[z]$ is the Hubble parameter.  Here the common $H_0=H[0]$ and
$h=H_0/(100\,{\rm km/s/Mpc})$ is used.

To compute the late-time peculiar velocity dispersion of the baryons
approximate by using the linear theory peculiar velocity dispersion given by 
\begin{eqnarray}
v_{\rm rms,lin}[z]&=&c\,\beta_0\,{H[z]\over H_0}\,{f[z]\,D[z]\over1+z} 
\\ \nonumber
\beta_0&=&{H_0\over c}
\sqrt{\int_0^\infty{dk\over k^3}\Delta^2_{\rm lin}[k,z=0]}
\end{eqnarray}
where $z$ is now the cosmological redshift, $D[z]$ is the linear theory
growth factor, $f[z]=-(1+z)D'[z]/D[z]\approx\Omega_\rmm[z]^{0.6}$, and
$\Delta^2_{\rm lin}[k,z]$ is the linear theory dimensionless power spectrum
(see \cite{p99} \S16.2).   For lo-$z$, $D[z]$ is approximately 
the solution to
\begin{equation}
D''[z]+{q[z]\over1+z}\,D'[z]-{3\over2}{\Omega_\rmm[z]\over(1+z)^2}\,D[z]=0
\end{equation}
with boundary conditions $D[0]=1$ and $D[\infty]=0$.  Here
$\Omega_\rmm[z]=8\pi G\rho_\rmm[z]/3 H[z]^2$ is the matter density parameter 
and $q[z]=-1+(1+z)\,H'[z]/H[z]$ is  the deceleration parameter.  In terms of a
sum over the different components, $c$,
\begin{equation} 
H[z]=H_0\,\sqrt{\sum_c \Omega_{c0}(1+z)^{3(1+w_c)}}
\end{equation}
where $\Omega_{c0}$ gives the present density parameter of each component and
$w_c$ the equation-of-state ($p/(\rho c^2)$).  In the concordance (flat
$\Lambda$-CDM) model $c$ sums over $\Lambda$ (for a cosmological constant), m
for matter (baryons + cold dark matter), r for radiation (photons and light
neutrinos).  In the assumed flat cosmology
$\Omega_{\rmm0}+\Omega_{\rmr0}+\Omega_{\Lambda0}=1$.  The equations of state
are $w_\Lambda=-1$, $w_\rmm=0$, and $w_\rmr={1\over3}$.

The power spectrum is approximated by 
\begin{eqnarray}
&&\Delta_{\rm lin}^2[k,0]=A\,k^{n+3}\,
T_{\rm BBKS}\left[{k\over h \Gamma}\right]^2
\\ \nonumber
&&\Gamma=\Omega_{\rmm0}\,
\exp\left[-\Omega_{\rmb0}\,\left(1+{\sqrt{2h}\over\Omega_{\rmm0}}\right)\right]
\\ \nonumber
&&T_{\rm BBKS}\left[q\right]={\ln[1+2.34\,q]\over 2.34\,q\,
\sqrt[4]{\hskip-10pt{1+3.89\,q+(16.1\,q)^2
         \atop\hskip20pt+(5.46\,q)^3+(6.71\,q)^4}}} \\ \nonumber
&&W_{\rm ball}[x]=3{j_1[x]\over x} \\ \nonumber
&&\sigma_8^2=\int_0^\infty{dk\over k}\Delta_{\rm lin}^2[k,0]
                       \,W_{\rm ball}[k\,8\,h^{-1}{\rm Mpc}]^2\ .
\end{eqnarray}
where $T_{\rm BBKS}$ is taken from \cite{BBKS}.  The last equation serves to
normalize $A$ in terms of the observational parameter $\sigma_8$.

Finally one needs the optical depth. The $z=0$ electron density is
$n_{\rme0}=3H_0^2\Omega_{\rmb0}/(8\pi G\mH)\,(1-{1\over2}Y_{\rm He})$ where
$\mH$ is the mass of the hydrogen atom and $Y_{\rm He}$ is the helium mass
fraction.  Thus
\begin{eqnarray} 
{d\overline{\tau}\over dt}&=&\tau_0\,H_0\,(1+z)^3\,\chi_\rme[z] \\ \nonumber
\tau_0&\equiv&\Omega_{\rmb0}{3\,c\,H_0\sigmaT\over8\pi G\mH}\,
                   \left(1-{1\over2}Y_{\rm He}\right)
\end{eqnarray}
where $\chi_\rme$ is the fraction of electrons which are ionized.  Here is
assumed instantaneous reionization at redshift $\zrei$ which is implicitly
defined by
\begin{equation}
\label{tauintegral}
\tauobs=\tau_0 \int_0^\zrei dz\,(1+z)^2\,{H_0\over H[z]}
\end{equation}
in terms of the observed optical depth, $\tauobs$.  Thus one finds that
\begin{equation}
\label{uvintegral}
\bar{u}_\rmv=u_{\rm 0}\,\int_0^\zrei dz\,{H[z]\over H_0}\,(f[z]\,D[z])^2
\end{equation}
where $u_0={2\over3}\tau_0\,\beta_0^2$.

To get numbers out the following numerical values for the cosmological
parameter: $h=0.73$, $\Omega_{\rmm0}=0.24$, $\Omega_{\rmb0}=0.0416$, $Y_{\rm
He}=0.24$, $\sigma_8=0.756$, and $\tauobs=0.9$ are chosen based on
ref.s~(\cite{SDSSlrg,WMAP}).  Thus $\tau_0=1.85\times10^{-3}$, $\zrei=10.3$,
$\beta_0=2.7\times10^{-3}$, $u_0=9.1\times10^{-9}$, and finally
\begin{equation}
\uv=3.4\times10^{-8}\ .
\end{equation}
This is very small!

To see how uncertainties in cosmological parameters can change this number, let
us first note that  the growth function is well approximated by
\begin{equation}
\label{ApproximateGrowth}
D[z]\approx{1\over1+z}\left({\Omega_\rmm[z]\over\Omega_{\rmm0}}\right)^{0.21}
\end{equation}
for $z\simlt20$.  Next note that the integrals of
eq.s~(\ref{tauintegral},\ref{uvintegral}) are dominated by $z\sim\zrei$ where
the cosmological constant is negligible and 
$\Omega_\rmm[z]\approx f[z]\approx1$ while
$H[z]\approx\Omega_{\rmm0}^{1/2}(1+z)^{3/2}H_0$ so 
\begin{eqnarray}
\tauobs&\approx&{2\over3}\,\tau_0\,\Omega_{\rmm0}^{-1/2}\,
(1+\zrei)^{3/2} \\
\nonumber
\bar{u}_\rmv  &\approx&{1\over3}\,\tau_0\,\beta_0^2\,\Omega_{\rmm0}^{0.08}\,
(\sqrt{1+\zrei}-1)
\end{eqnarray}
or eliminating $\zrei$ one finds
\begin{equation}
\bar{u}_\rmv\approx{1\over3}\beta_0^2\,\tau_0\,\Omega_{\rmm0}^{0.08}\,\left(
\left({3\over2}\sqrt{\Omega_{\rmm0}}{\tauobs\over\tau_0}\right)^{1\over3}-1
\right)\ .
\end{equation}
One sees a relatively weak dependence on $\Omega_{\rmm0}$ and $\tauobs$; while
$\tau_0\propto\Omega_{\rmb0} H_0$ is fairly well constrained.  By far the
largest uncertainty is through $\beta_0^2\propto\sigma_8^2$.  There is
considerably debate as to the value of $\sigma_8$.  There are also non-linear
corrections to the rms velocity which have not been included.

Ref.~\cite{ZhangPenTrac} have given a heuristic formula for the late-time tSZ
effect which includes only heating from gravitational collapse is
\begin{equation}
\ySZ[t_0]=1.5\times 10^{-6}\,
\left({\sigma_8\over0.756}\right)^{4.1-\Omega_{\rmm0}}
\left({\Omega_{\rmm0}\over0.24}\right)^{1.26-\sigma_8}\ .
\end{equation}
For the choice of cosmological parameters given above one finds that the tSZ
$y$-distortion is $\sim75$ times larger than that from scattering of
anisotropies.

\end{document}